\documentclass[12pt,letterpaper]{exam}

\usepackage{graphicx}        
\usepackage{multicol}        
\usepackage[bottom]{footmisc}

\usepackage{newtxmath}       

\usepackage{natbib}

\usepackage[usenames,dvipsnames,svgnames,table]{xcolor}
\usepackage[margin=1.0in]{geometry}

\usepackage{color}
\usepackage{epsfig}
\usepackage{hyperref}
\usepackage{amsmath,amssymb}
\usepackage[shortlabels]{enumitem}
\graphicspath{{./img/}}

\newcommand{\FF}{\mathcal{F}} 
\newcommand{\Xbf}{\mathbf{X}}
\newcommand{\Sbf}{\mathbf{S}}

\begin{document}

\title{Vegetation Pattern Formation in Drylands}
\author{ Punit Gandhi
\thanks{Mathematical Biosciences Institute, Ohio State University, Columbus, OH 43210}, 
Sarah Iams
\thanks{Paulson School of Engineering and Applied Sciences, Harvard University, Cambridge, MA 02138}, 
Sara Bonetti
\thanks{Soil and Terrestrial Environmental Physics, 
ETH Zurich, Zurich, Switzerland} and 
Mary Silber\thanks{
Department of Statistics, University of Chicago, Chicago IL 60637 email: msilber@uchicago.edu}
}

%
%
\maketitle
\abstract{This is a book chapter, written as a contribution to a new edition of  \textit{Dryland Ecohydrology}, edited by Paolo D$'$Odorico, Amilcare Porporato, and Christiane Runyan, (to appear, Springer 2019). It aims to (1) describe some of the background to conceptual mathematical models of spontaneous pattern formation, in the context of dryland vegetation patterns, and (2) review some of the observational studies of the phenomenon. The chapter also highlights challenges and opportunities associated with the development of the models in light of increasing availability of remote sensing data. This includes both satellite imagery of the patterns and elevation data of the topography.  The vast scales, in time and space, associated with the key processes further suggest avenues for improved mathematical modeling paradigms. 
}

\section{Introduction}
\label{sec:introduction}

\begin{figure}
\begin{center}
\includegraphics[width=\linewidth]{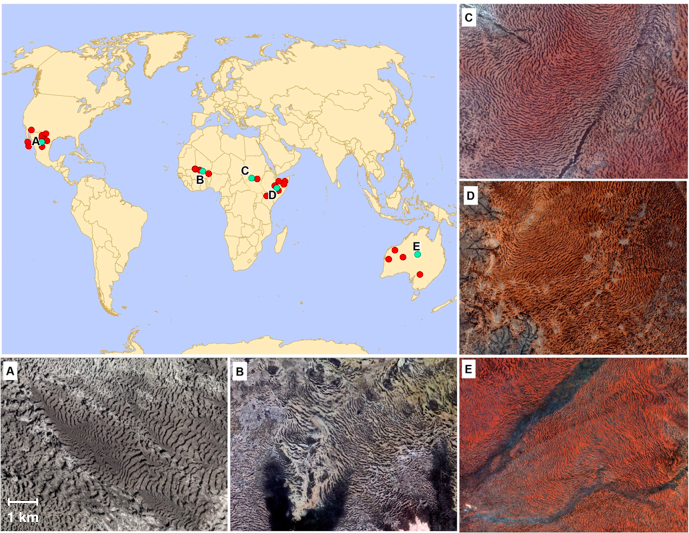}
\end{center}
\caption{Global distribution of vegetation bands with examples from three continents. Red dots represent banded locations identified by 
\cite{deblauwe2012determinants} or by Jake Ramthun. (A) Sierra del Diablo, Mexico (27.93$^\circ$ N, 103.35$^\circ$ W, 5/11/2016). (B) Mopti Region, Mali (14.64$^\circ$ N, 1.56$^\circ$ W, 3/6/2016). (C) Muglad, Sudan (11.27$^\circ$ N, 28.23$^\circ$ E, 3/20/2016). (D) Gode, Ethiopia (5.95$^\circ$ N, 43.96$^\circ$ E, 3/19/2016). (E) MacDonnell Range, Australia (23.39$^\circ$ S, 133.86$^\circ$ E, 3/20/2016). All five images are on the same scale.  Figure courtesy of Jake Ramthun.}
\label{fig:jake}
\end{figure}

Some of the most beautiful examples of spontaneous pattern formation occur in drylands, where the spatial distribution of vegetation may appear as a remarkably self-organized patchiness on a scale spanning kilometers. Figure~\ref{fig:jake} shows examples of arced bands of vegetation, which alternate  rhythmically with bare soil, in different drylands around the globe. Such examples of community-scale self-organization of water-limited ecosystems were first reported when aerial photographs accompanied hydrological survey work in the Horn of Africa, dating to the 1940s. The aerial surveys revealed the extent of these large-scale patterns, something that was not possible from the ground. 
Modern satellites have since provided a valuable tool for investigating these striking natural patterns. 

A sense for ``pattern" is well-honed in humans, and yet the term can lead to different interpretations or definitions. In this chapter, we restrict its scope to a mathematical one of possessing spatial symmetries. 
An idealization of vegetation bands, such as those shown in  Figure~\ref{fig:jake}, exhibit translation symmetry that is continuous along the bands.  They also exhibit discrete translation symmetry, transverse to the bands; the pattern would look about the same if we translated everything  by one stripe.  In another type of vegetation pattern, a regular pattern of bare patches appears in an otherwise vegetated region.  These types of patterns possess discrete rotation and translation symmetries, where rotating the pattern by a certain amount or translating it by a certain amount brings us to the same pattern.
We focus on patterns that  emerge in response to aridity stress, where patchiness may represent a strategy for harvesting the limited water from a region of suitable scale to sustain the patch. We would like to understand what sets the scale of the patches and the symmetries associated with their distribution.  Even starting from knowledge of the biomass coverage fraction, these are not obviously determined.  Instead, to explain the distribution of biomass, it's necessary to understand the cooperative feedback between plants and nutrients, as well as the mechanisms for spatial redistribution of the limiting resources. 

Modeling natural spatial patterns emerged as a challenging test bed for theories of pattern formation after a boom of nonlinear pattern formation research occurred in the 1980s based in laboratory sciences.  This chapter highlights examples of both observational studies and mathematical modeling ones.
Table~\ref{tab:wetvsdry} summarizes a number of the challenges that arise in modeling of the vegetation pattern formation process, including a lack of fundamental equations and a lack of controlled experiments.  In addition, there is an enormous range of  potentially relevant temporal and spatial scales. Timescales may vary from those associated with rain events (minutes, hours) to those associated with plant colonization and regeneration (decades, centuries). Likewise spatial scales range between those of individual plants (centimeters, meters), to those of the macroscopic community-scale patterns and landscape (decameters, kilometers). Often models are constructed by considering the most relevant processes, which may act on vastly different spatial and temporal scales, and incorporating them in the model on the time and length scale of the self-organized vegetation patches. These conceptually simple models are amenable to mathematical analysis. They are intrinsically limited, however, because they do not resolve processes on the appropriate scales, nor are they developed through explicitly averaging equations over finer scales to develop the simplified model.  Because of this, it can be challenging to constrain parameters, or even to identify how the parameters correspond to measurable quantities. These issues are described through examples in Section~\ref{sec:models:examples}. 

\begin{table}
\begin{center}
\begin{small}
\begin{tabular}{|c|c|c|}
\hline
&Wetlands&Drylands\cr
&(Pattern Formation in Fluid Dynamics)& (Pattern Formation in the Environment)\cr
\hline
Equations&Navier-Stokes&models exist\cr
\hline
Parameters&often excellent specs&some known, order of magnitude level\cr
\hline
Time-Scales&$\ll$PhD thesis scale&decades-century\cr
\hline
Space-Scales&table-top&meters to landscape-scale\cr
\hline
Symmetries&excellent approximation&heterogeneities offer opportunities?\cr
\hline
Mechanisms&well developed and validated&generic mechanisms invoked\cr
\hline
\end{tabular}
\end{small}
\end{center}
\label{tab:wetvsdry}
\caption{Contrasting the challenges of modeling pattern formation of vegetation patterns with those in the setting of a typical physical system where the theory was developed.}
\end{table}%

The conceptual approach, in contrast to detailed mechanistic modeling, typically exploits our mathematical understanding of generic mechanisms for pattern formation. This perspective is at the heart of proposals for using pattern characteristics as early warning signs of ecosystem collapse, as described in Section~\ref{sec:models:warning}. 
The expectation is that the pattern formation phenomena may rely on few details of the system beyond which spatial symmetries are present.  This, afterall, explains why similar patterns are observed in such disparate physical, chemical and biological systems. In this instance, it can also explain why similar patterns appear in such far-flung regions of the globe, and in different formulations of the mathematical models of vegetation patterns. However, generic arguments often only apply in ``a neighborhood" of an instability and if the patterns are of ``sufficiently small" amplitude. Moreover, certain mathematical assumptions must be satisfied by the governing equations for these generic arguments to hold. These assumptions may be built into a conceptual mathematical model, but whether the assumptions hold cannot be checked against fundamental equations for the system (fundamental equations would be akin to Navier-Stokes for fluids), because such fundamental equations do not exist. This leads naturally to caveats on model predictions that are often unstated. These limitations on the predictions of models are important to make explicit in the context of predicting the fate of an ecosystem under changes in the degree of environmental stress, or in human land use.

This chapter focuses on observations and conceptual mathematical modeling of vegetation patterns that exhibit a dominant spatial scale and self-organization into a regular pattern. It omits other fascinating threads of dryland pattern formation research where there is no dominant scale, e.g. scale-free patch-size distributions that have been investigated in arid Mediterranean ecosystems~\citep{kefi2007spatial} and, more globally, in the context of dryland ecosystem multifunctionality~\citep{Berdugo2017}. Nor does it review literature on multiple spatial scale vegetation patterning, such as has been investigated in the context of so-called ``fairy circles"~\citep{Bonachela2015,getzin2016discovery,Tarnita2017}. 
The perspective is admittedly skewed towards the pattern formation one that emerged from investigating physical systems.
\cite{meron2012pattern,meron2015nonlinear,meron2018patterns} provides further background on the physics approach to pattern formation, specifically in the context of ecological modeling. 
\cite{Borgogno2009} is an excellent review article on the mathematical modeling frameworks for vegetation patterns. An overview of large scale regular pattern formation  in a variety of ecosystems, including mussel beds, wetlands, drylands, ribbon forests and mudflats, is provided by \cite{rietkerk2008regular}.

This chapter is organized as follows.  Section~\ref{sec:observations} reviews observational studies of vegetation patterns, starting with the early ground-based and aerial-survey ones from the Horn of Africa, and moving on to modern observational studies that are informed by mathematical models and often incorporate satellite data. Section~\ref{sec:observations} ends with a summary of the key ecohydrological processes that are implicated in the pattern formation process. Section~\ref{sec:models} introduces some of the mathematical theory of pattern formation that informs studies based in model simulations. It 
describes mechanisms of vegetation pattern formation, emphasizing the way that our understanding of pattern formation in physical and chemical systems has  influenced model development. It also presents a sequence of three conceptual mathematical models that frame the pattern formation problem, each of which adds a layer of complexity on its predecessor. Section~\ref{sec:models} ends with a discussion of the challenges of constraining parameters in such parsimonious models.
Section~\ref{timescales} turns to the modeling challenges that arise because many of the key processes act on vastly different time and space scales. It identifies some possible mathematical directions that might turn this challenge into an opportunity, both for improved modeling, by better constraining parameters, and for speeding up simulations by separating the ``fast" from the ``slow" scales. Section~\ref{timescales} ends by highlighting two specific directions as promising: modeling  stochasticity of rain events explicitly, and exploiting landscape scale topographic heterogeneity to probe models.
The chapter closes in Section~\ref{sec:outlook} with our outlook, which emphasizes the importance of including humans in the equations.

\section{Observational Studies}
\label{sec:observations}

\begin{figure}
\begin{center}
\includegraphics[width=\linewidth]{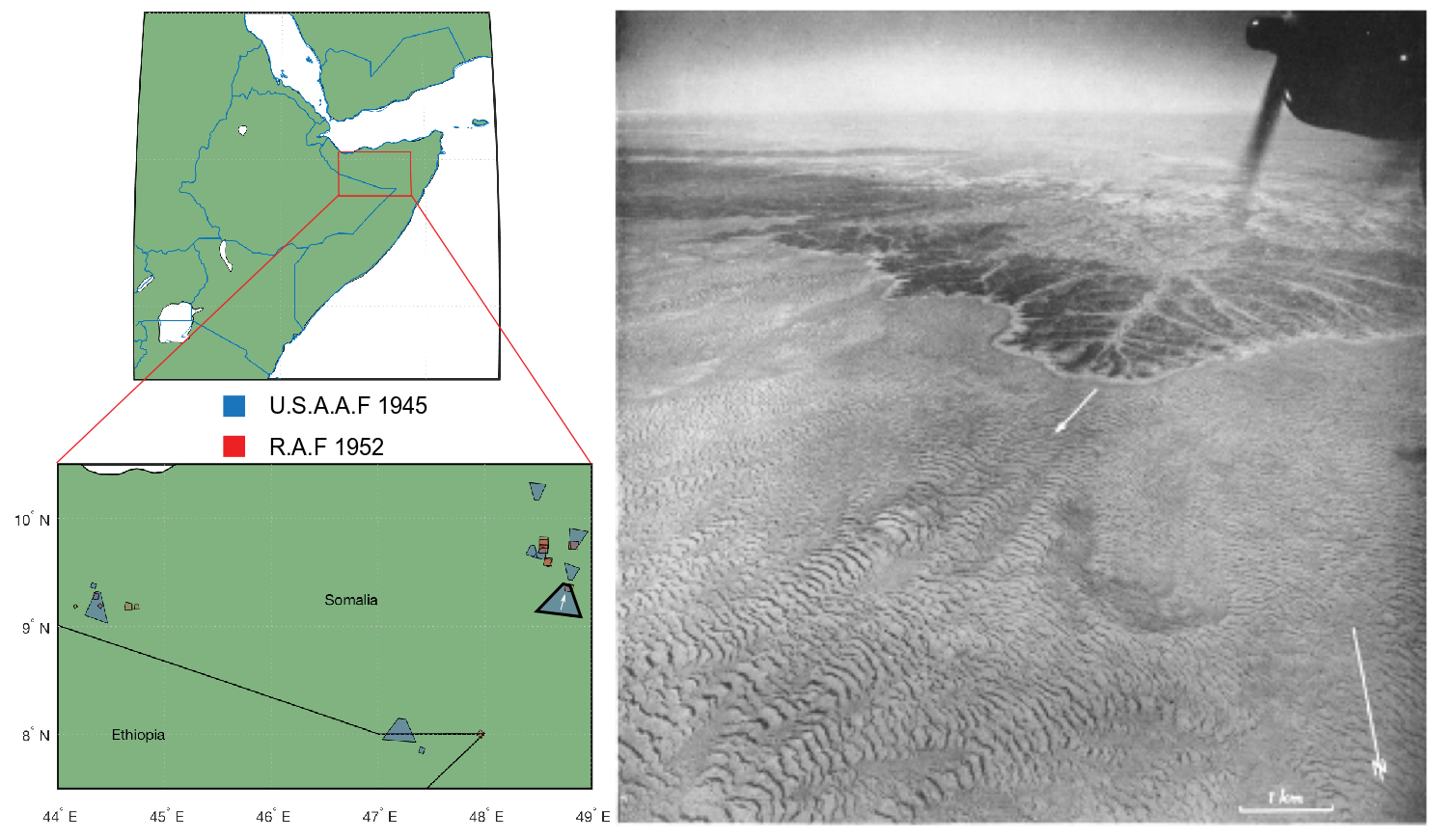}
\end{center}
\caption{ The locations of aerial images take during a March 1945 US Army Air Force reconnaissance mission that are reproduced in 
\cite{macfadyen1950vegetation} are shown in blue in the lower left map.  Those from the Royal Air Force taken in 1950 and reproduced in
\cite{greenwood1957development} are shown in red.  In the photograph on the right, corresponding to the region outlined in bold on the map, the dark stripes are  vegetation bands and the light regions in between are bare soil.
Figure courtesy of Karna Gowda.}
\label{fig:aerial}
\end{figure}
We look at some of the observational work on vegetation arcs, and other regular vegetation patterns, and  summarize the key ecohydrological processes involved.  The observational studies include aerial surveys and satellite image analysis, as well as some ground-based field investigations. Unlike early studies (circa 1950 to 1995), more recent investigations (1995 to present) are informed, in part, by models. 

\subsection{Early Investigations}
\label{sec:earlyinvestigations}
Imagery of banded vegetation patterns dates to the first half of the 20th century.  In the late 1940s, W.A.~Macfadyen, a geologist working on a water supply investigation in what was then British Somaliland \citep{macfadyen1933geology}, identified banded vegetation patterns on aerial photographs and reported them in a letter to Nature \citep{macfadyen1950soil}.  He soon published a more extensive description in the Geographical Journal~\citep{macfadyen1950vegetation} where he reproduced aerial imagery of the bands taken in 1945 by the US Army Air Force
(U.S.A.A.F.).  Although there had been some aerial photographs of Somaliland dating to 1920~\citep{macfadyen1931taleh} and an image of a single band appeared in a 1939 book focusing on soil erosion \citep{jacks1939rape}, patterns had not before been reported.  Macfadyen's letter to Nature \citep{macfadyen1950soil} generated additional reports of patterns in Sudan and in the Kalahari region.  

Figure~\ref{fig:aerial} shows one of the images reproduced in Macfadyen's paper.  The geographic locations of those images, after cropping any horizon lines, are represented by the blue boxes on the map. 
In addition to working with aerial imagery, Macfadyen surveyed a number of bands.  He measured the local slope of the ground where arcs occurred ($0.2$ to $0.3$ \% grades), the width of the vegetation bands ($10$ to $76$ m) and the wavelength (or spacing) of the bands ($70$ to $276$ m) \citep{macfadyen1950vegetation}. He noted that the arcs were oriented convex upslope in his region of study, thereby providing an indicator, from the aerial images, of the direction of a gentle prevailing grade. (One of his drawings of the arcing relative to the hillslope is reproduced in Figure\ref{fig:arcing}a.)  In addition to characterizing the bands, he reported on the hydrology, linking it to the  arcs. In a region near the bands, Macfadyen described rare, large rainstorms driving a film or sheet of water flowing overland (up to 25 mm deep).  Macfadyen thought the relevant storm frequency might be as low as one or two such storms every few years.  He speculated that overland water transport of organic matter, including seeds and animal excrement, by such rare storms might account for the origin of the arcs. 

Macfadyen's observations in the Horn of Africa were followed by those of Greenwood, in the same region~\citep{greenwood1957development}.  Greenwood focused on Royal Air Force (R.A.F.) aerial images dating to a 1951-52 aerial survey.  The locations of images he reproduced in his paper are shown, in red, in Figure~\ref{fig:aerial}.
The R.A.F. aerial survey images remain available via the Bodleian Library at the University of Oxford~\citep{gowda2018signatures}.  Making a site visit in 1955-56, Greenwood examined soil composition and made topographic observations.  He linked water availability for vegetation to the inverse of soil clay content, noting that in Sudan, higher clay content soils require more rain to support trees than do less clayey soils.  Water availability is currently an important component of modeling efforts, but the role of soil composition is hard to quantity in large scale studies of patterns due to dataset limitations \citep{bastiaansen2018multistability}.

Greenwood's topographic observations extended Macfadyen's report that arcs curve convex up-slope.  Greenwood noted that they curve up-slope in shallow valleys, can be virtually straight on slight slopes, and can be convex down-slope when on low ridges.  Arc curvature is discussed in Section~\ref{sec:topography} in the context of current mathematical modeling efforts.
Greenwood also provided a mechanistic description of how bands might form.  He identified vegetative litter mulch as creating a trio of positive feedbacks: it would increase penetration of water into the soil, reduce evaporation from the soil surface, and trap further organic litter from run-on.  These are still believed to be important feedbacks in the system: feedback effects associated with increased infiltration and reduced evaporation have been incorporated into contemporary models \citep{rietkerk2002self,gilad2004ecosystem} and a feedback related to the transport of surface organic matter is also sometimes included \citep{saco2007eco}.

From Sudan, J.W. Wright reported grass patterns to Macfadyen in 1950 \citep{macfadyen1950vegetation}. In 1959, G.A. Worrall published accounts of those grass patterns \citep{worrall1959butana} and of tree patterns, in a region with sandier soil, in 1960 \citep{worrall1960tree}.  In the grassy region, stripes were on high ground in between rivers, and arced convex downslope.   Here the patterns sit on land with slopes between $0.2\%$ and $0.5\%$ with $8$-$12$ meters of vegetation separated by twice that width of bare ground.  Worrall finds a substantial moisture differential between vegetated and bare ground.  He observed rain led to moisture penetration of 10-15 cm in bare areas and 30-40 cm under grass, and that vegetation follows moisture, with lusher upslope growth, leading to upslope bands movement. Water transport was impacted by the formation of potholes at the upslope frontal edge of the stripes as well as by cracking of the soil within the stripe.  Worrall identified feedbacks between water, wind, topography, soil, and vegetation as all contributing to the patchy biomass distribution associated with vegetation patterns \citep{worrall1960patchiness}.

After these initial works establishing the phenomena, observational efforts continued through the 1960s.  Boaler and Hodge worked with patterns of grasses, shrubs, and Acacia in the Horn of Africa occurring on slopes of $0.2\%$ to $0.7\%$ \citep{boaler1964observations}, and also noted soil potholes impacting water transport at the front and in the body zone of the arcs.  Observing a rainstorm, they found infiltration in the vegetated zone was five times deeper than in the bare region between arcs.  Boaler and Hodge also speculated about an arc formation mechanism that starts with uniform vegetation under wet (or low grazing) conditions, with bare patches arising as the climate dries (or grazing increases).  They imagine that plants adjacent to open spaces might be at a disadvantage, until there is a sufficient bare area for substantial runoff of water, creating a selection mechanism for the size of bare areas.

\begin{figure}
    \centering
\includegraphics[width=0.9\linewidth]{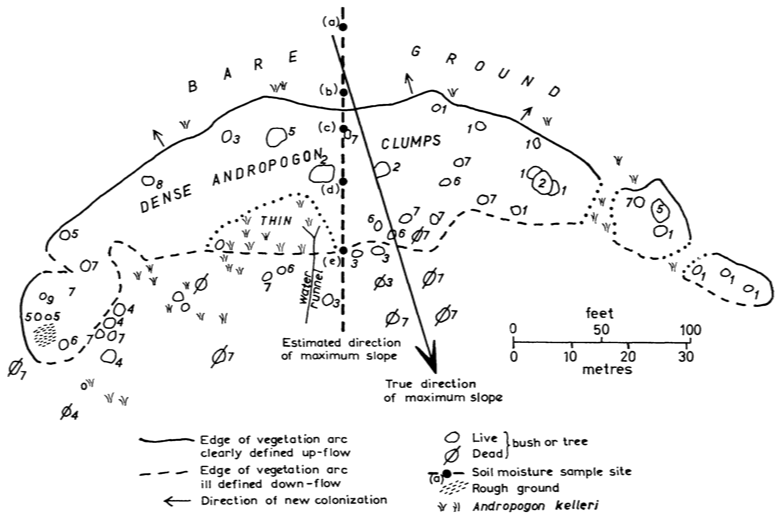}
    \caption{Detailed drawing of a vegetation arc in Somalia, reproduced from 
    \cite{hemming1965vegetation}.}
    \label{fig:Hemming}
\end{figure}

Hemming also thought about the arcs as being in balance with water availability via runoff and grazing pressure \citep{hemming1965vegetation}.  He thought increasing rainfall or decreasing grazing might widen individual arcs, while decreasing rainfall might reduce the number of arcs.    
In addition to thinking about how these systems stay in balance, Hemming contributed careful observational work.  A map he constructed of an arc near Las Anod in northern Somalia is reproduced in Figure~\ref{fig:Hemming}.  He catalogued vegetation \citep{hemming1966vegetation}, observed runoff, and measured soil saturation.  For example, in one observation of sandy soil (8.64N, 47.58E) with 36 mm of gentle rain, the top 60 cm of soil saturated before runoff occurred.  He also made an estimate of 15-30 cm/year migration speed, via upslope colonization, for his surveyed arc.  This number is in the range of some contemporary measurements \citep{bastiaansen2018multistability,deblauwe2012determinants,gowda2018signatures}.

By 1970 striped patterns had been reported in the Somali Republic, the Republic of Sudan, Niger, Australia, Saudi Arabia and Iraq \citep{wickens1971some}.  A distinction had been drawn between these stripes, that seem to form from a water-vegetation interaction, and those in Utah in the United States,  Jordan, and Mauritania, where striping may be more related to wind-blown material \citep{white1971vegetation}.  Theories of stripe formation via water-vegetation interaction were summarized by  
\cite{white1971vegetation}.  He expanded upon the ideas of Boaler, Hodge, and Hemming by thinking about vegetation arcs across a precipitation gradient.  He imagined that stripes might form via bare spots developing in uniform vegetation at the wetter end of the zone, and from colonization of largely bare surfaces at its arid margin.  Greig-Smith's 1979 presidential address to the British Ecological Society summarized the work of \cite{boaler1964observations}, \cite{hemming1965vegetation} and \cite{white1971vegetation} 
on vegetation arcs and theories of their formation, while also providing a much broader context for non-randomness of vegetation distribution across many ecosystems \citep{greig1979pattern}.

By the 1970s, a list of factors that need to be present for vegetation stripes to occur was also well developed.
\cite{wickens1971some} offer a list of factors  that is similar to a list offered by 
\cite{boaler1964observations} a few years before.  They name the following factors: patterns were known to occur in warm arid and semiarid regions on a variety of soils; patterns had been seen at rainfalls of between $100$ and $500$ mm per year, with rain often coming in heavy events; surface slopes ranged from $0.2$\% to $2$\%; shallow enough slopes that water runs off in sheets, rather than draining via channels; and across all of these locations, bare ground has a low infiltration rate (even with sandy soils, due to surface crust), while vegetated locations accumulate water.  

In the 1980s and 1990s, case studies were reported from Mexico, extensive research was ongoing in Australia, and a large-scale study of the Sahel was undertaken in Niger.  The work of Cornet and Monta\~na in the Chihuahuan desert of Mexico \citep{cornet1988dynamics,montana1990response,montana1992colonization} adds observations of soil moisture, plant migration, and plant community evolution drawn from nearly a decade of study.  Their observations at the transition zone where striped vegetation gives way also drove a hypothesis about landscape evolution.  They posited that the water drainage system gave way from gully-flow to sheet flow, leading to the development of vegetation stripes \citep{cornet1992water}.  In Australia, work by 
\cite{ludwig1995spatial,tongway1990vegetation} followed up on the early research of \cite{slatyer1961methodology}
on \emph{Acacia aneura} grove/intergrove patterns.  They identified microtopography \citep{tongway1990vegetation}, developed monitoring methods \citep{tongway1995monitoring}, and also made observations of patch size distribution and its sensitivity to rainfall \citep{ludwig1999fine}.  

In Niger, the HAPEX-Sahel study \citep{goutorbe1997overview} included a ``Tiger bush'' (banded vegetation) study site.  The February 1997 issue of the Journal of Hydrology was a special issue reporting on the study.  The issue included detailed rainfall information \citep{lebel1997rainfall}, evaporation measurements  \citep{wallace1997soil, kabat1997evaporation} and carbon dioxide flux \citep{levy1997co2} in the banded vegetation patterns, as well as evapotranspiration measurements at a savannah site nearby \citep{tuzet1997flux}.  A September 1999 special issue of the journal Catena includes additional case studies from Niger \citep{chappell1999testing,galle1999water,valentin1999soil}, Mexico \citep{janeau1999soil}, and Australia \citep{dunkerley1999banded,macdonald1999distribution} addressing water transport, water harvesting, band migration, and soil cations distribution.  In addition, a more recent paper documents an ecohydrological study of a patterned site in the Stockton Plateau of West Texas in the United States \citep{mcdonald2009ecohydrological}.

\subsection{Recent observational work}

Due to advances in model development (see Section \ref{sec:models:development}) that occurred in the late 1990s, more recent observational work has often been influenced by modeling efforts. 
\cite{barbier2014case} have reviewed a number of recent observational results in the context of model predictions.  Starting with the work by
\cite{barbier2006self}, observations are often presented in the context of model predictions, and make use of the multidecadal aerial images to identify changes in vegetation (e.g., in Niger over a forty year interval).  The use of multiple images over time was pioneered in work by 
\cite{wu2000fragmentation}, where they took advantage of three decades of aerial photographs in Niger to examine fragmentation.  
\cite{barbier2006self} found that extended drought in their study region in Niger was accompanied by a shift from homogeneous vegetation to spotted patterns, and that the shift was lessened in areas protected from wood cutting and grazing. They connected their results to modeling efforts by noting that the observed spatial wavelength of the patterning appeared to be endogenous.  They saw this as offering support for models that rely on a pattern formation mechanism to generate patterns (see Section \ref{sec:models:mechanisms}.) 

\begin{figure}
    \centering
      \includegraphics[width=\linewidth]{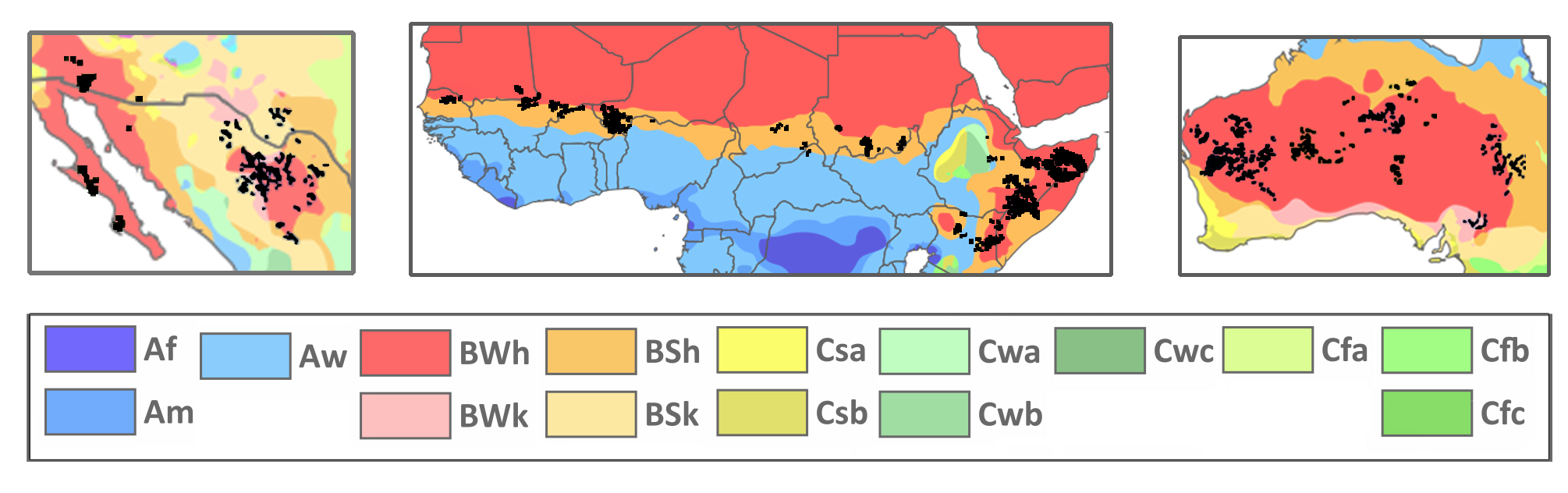}
    \caption{ Patterned sites in (left) North America, (center) Africa and (right) Australia are shown in black from~\cite{deblauwe2008global}. Maps are colored based on the K\"oppen--Geiger climate classification system~\citep{peel2007updated}, where the first letter labels  Tropical (A) , Arid (B) and Temperate (C) climates. BWh and BSh are warm arid and warm semiarid climates,  the one most associated with patterns.  
    }
    \label{fig:deblauwe}
\end{figure}

Recent observational work has often made use of remotely sensed data, and sometimes of contemporary statistical learning methods.  Some new pattern locations were identified by 
\cite{deblauwe2008global} after studying correlates of patterning using a MaxEnt machine learning algorithm developed for identifying factors of ecological niche segregation \citep{elith2011statistical}.  Figure \ref{fig:deblauwe} shows some of the pattern locations in North America, northern Africa, and Australia, where they are in black over a map based on the K\"oppen-Geiger climate classification system, indicating that patterns are mainly observed in warm arid (BWh) or warm semiarid (BSh) regions.  

There has also been some work oriented towards directly constraining model parameters \citep{barbier2008spatial}, however, many recent studies focus on measuring quantities that are common outputs of a model, such as pattern morphology or pattern wavelength, as a function of precipitation or other environmental factors \citep{deblauwe2012determinants,deblauwe2011environmental, penny2013local}.  Pattern morphologies include homogeneous vegetation cover, vegetation with bare patches (gaps), labyrinthine patterns, isolated patches or spots of vegetation, and bare ground.  In models, different morphologies are associated with different levels of mean annual precipitation, and if precipitation levels were to change with time, transitions between pattern states is often predicted.  See Section \ref{sec:models:warning} for more about pattern morphology and critical transitions. 

In a study of Sudan by 
\cite{deblauwe2011environmental}, they analyze satellite imagery from 1967, 1988, and 2001-2002 using Fourier-based methods to identify pattern morphology as a function of the rainfall gradient across the study region (approx. 350mm/yr to 500 mm/yr). Figure \ref{fig:collapse} in Section \ref{sec:models:warning} is reproduced from their work, and indicates model predictions of pattern morphology in an aridity-slope parameter space.  Their observations qualitatively matched the predictions of how patterns would be distributed along a rainfall gradient.  In addition, the time interval in their study included decades of low precipitation.  
The change in precipitation coincided with regions that transition from gapped patterns to labyrinths as well as regions that transition from labyrinths to spotted patterns.
This type of morphological change with reduced precipitation aligns with model predictions.  In addition to comparing pattern morphology, features such as the wavelength and orientation of bands relative to the local slope \citep{bastiaansen2018multistability,penny2013local}, the width of bands \citep{gowda2018signatures}, the local biomass \citep{bastiaansen2018multistability}, and arc curvature \citep{gandhi2018topographic}, have been identified using satellite or aerial imagery and compared to model predictions. 
Other work has focused on vegetation patch size distribution \citep{kefi2007spatial} as an indicator of desertification \citep{Berdugo2017, genin2018spatially,moreno2011assessing}.  

\subsection{Summary of Key Ecohydrological Processes}
\label{sec:ecohydrology}

Following the basic pattern formation concepts, often attributed to Alan Turing, vegetation patterning is thought to be the result of an interplay between short-range cooperative and long-range
competitive plant interactions  \citep{Borgogno2009,lejeune2004vegetation,rietkerk2008regular}. 
Here we summarize some of these feedbacks, many of which have been incorporated in mathematical models of vegetation pattern formation. 

Positive feedbacks between plants and water include the ability of plants to concentrate resources (e.g., runoff interception, shadow, nutrient-rich litter) providing more favorable conditions for plants to establish. In particular, vegetation cover intercepts runoff thus favouring infiltration and increasing local soil water availability \citep{meron2015nonlinear,paschalis2016matching,saco2007eco,valentin1999soil,ursino2005influence}, and protects against soil erosion \citep{ludwig2005vegetation,saco2007eco,valentin1999soil}. In addition to this runoff-runon mechanism, infiltration rates under vegetated patches are enhanced as a consequence of the increased soil macroporosity related to biological activity (e.g., termites and earthworms) and vegetation roots \citep{saco2007eco}. In contrast,  bare areas are maintained by formation of  soil crusts that reduce water infiltration, preventing vegetation from establishing and growing \citep{assouline2015dual,Borgogno2009,meron2015nonlinear}.
In addition to this positive infiltration feedback, vegetation cover may decrease temperature fluctuations, exposure to solar radiation, and wind desiccation, thus reducing water losses and further favouring plant growth. Another positive feedback mechanism is the protection from fires and grazing \citep{Borgogno2009}. The concentration of resources (i.e., water and nutrients) in vegetated and fertile soil patches -- contrasted to unfertile bare areas -- has led to the concept of ``fertility islands'' \citep{ridolfi2008fertility,schlesinger1990biological}. Species that are capable of modifying the abiotic environment by resource redistribution, and thus facilitating the growth of other species, are known as ``ecosystem engineers'' \citep{gilad2004ecosystem,jones1994organisms}.
Conversely, competition for resources among plants is exerted thorough the root system which extends beyond the canopy area and extracts water and nutrients from the bare patches \citep{meron2016pattern,rietkerk2008regular}. This negative feedback takes place over a range that is  larger than that of facilitative interactions. 

In addition to these positive and negative feedbacks, a number of exogenous climatological and geomorphological factors can further affect vegetation patterning. In fact, spatial heterogeneities in soil properties and topographic features are likely to alter water runoff, infiltration, solar radiation and erosion processes \citep{paschalis2016matching}, ultimately changing the spatial redistribution of resources and thus the pattern formation mechanism. A number of studies empirically analyzed the relation between pattern geometry and ``external" environmental conditions, such as rainfall, temperature, ground slope, soil type and other topographic variables \citep{augustine2003spatial,deblauwe2008global,penny2013local}. Analysis of the location of vegetation patterns (i.e., their ``ecological niche") provided evidence for a correlation between vegetation patterns and environmental constraints (namely, humidity index, mean temperature and its seasonality, rainfall amount and seasonality, soil wilting point and ground slope) at global scale \citep{deblauwe2008global}.  Pattern morphology also seems to be highly related to large scale properties of the landscape such as slope orientation, hillslope gradient, and soil type, as well as smaller scale features (e.g., ridges, streams, roads, and soil peculiarities), which can produce deviations from the global trends \citep{penny2013local}.

\section{Nonlinear Dynamics Perspective: Critical Transitions and Pattern Formation}\label{sec:models}

The classic study by~\cite{pearson1993complex} of pattern formation in the Gray--Scott model~\citep{gray1983autocatalytic}  
demonstrates the rich and often unexpected pattern forming behavior by even the simplest of reaction-diffusion systems.    It is therefore not surprising that models of this type can reproduce observed spatiotemporal patterns. 
Section~\ref{sec:models:patternformation} discusses the development of the mathematical framework within nonlinear dynamics that connects the emergence of patterns across a broad range of contexts. 
The key idea is that as a parameter changes a uniform (vegetated) state loses stability and a patterned state emerges. 
The usefulness of this approach lies in the fact that the predictions are often insensitive to details of a given model, at least for parameters near where the patterns form. This is described further in the context of early warning sign proposals in Section~\ref{sec:models:warning}.  Section~\ref{sec:models:mechanisms} provides more details of the mathematical mechanisms behind ``pattern forming instabilities" in vegetation models, with a chronology of their development in Section~\ref{sec:models:development}. 
We also include some details of three particularly influential phenomenological models in Section~\ref{sec:models:examples}, and conclude in Section~\ref{sec:models:parameters} by illustrating  the challenges faced when trying to constrain parameters in these models.

\subsection{Pattern Formation}
\label{sec:models:patternformation}

The study of spontaneous spatial pattern formation took off as a sub-field of nonlinear dynamics research in the 1980s, although it has a much longer history within fluid dynamics research. See \cite{cross1993pattern} for a general review from physics, and  \cite{crawfordknoblochreview} for a review in the fluid dynamics setting. 
The nonlinear dynamics perspective emphasizes ``generic mechanisms". For example, those of bifurcation theory, which is a mathematical framework that helps  elucidate  the origin of bistability, hysteresis, and spontaneous oscillations in nonlinear systems. Studies in nonlinear dynamics  helped to identify fundamental routes to chaos, and also  the transitions to patterns in spatially extended nonlinear systems. 
With a focus on generic mechanisms, 
connections are  made between strikingly similar emergent behavior in chemical, biological and physical systems. 

At the heart of pattern formation research is a basic notion of ``spontaneous symmetry-breaking", which is associated with the onset of an instability. Consequently, genericity arguments apply for special parameter sets that are close to those where an instability sets in.
These arguments do not require much detailed knowledge about the system; just knowledge about the type of instability (e.g. whether it is oscillatory or steady/monotonic), and of the associated spatial symmetries of the system, and then one can  predict what kinds of patterns are possible~\citep{golubitsky2012singularities,hoyle2006pattern}.
While the stability of these patterns depends on details of the model nonlinearities, the characteristic spatial and temporal scales are often determined, approximately, by the quantitative linear computations that are required to determine the modes of instability.
However, the results are technically only valid for ``small amplitude patterns" near the onset of instability in parameter space. Some care must be taken when extending the results to the states actually observed in the environment, which go between well-developed community-scale vegetation patches and bare soil with zero biomass, and hence are not ``small amplitude" patterns.  

Much of the research on pattern formation in the 1980's and 90's involved a healthy back and forth between laboratory experiments and theory. An interesting counter point to this, which is relevant to vegetation pattern modeling, involves the so-called ``Turing mechanism". This mechanism is attributed to a mathematical argument of Alan Turing, which appeared in his 1952  paper ``The Chemical Basis of Morphogenesis"~\citep{turing1952chemical}.  In it, he shows that diffusing and reacting chemical species, which he calls ``morphogens", may undergo an instability that leads to a spatial pattern, i.e., the diffusing chemicals do not evolve to have spatially uniform concentrations. An essential ingredient for this pattern formation process is that the two species diffuse at drastically different rates: an activator chemical diffuses slowly, acting only at short-range, while an inhibitor chemical diffuses rapidly, and acts at long-range. The formation of a pattern with an emergent characteristic lengthscale is, on the surface, quite surprising  because it relies on diffusion, which we generally associate with destroying any sharp boundaries, such as those that appear in the patterns. The emergent pattern lengthscale is tied to the competition between two drastically different diffusive lengthscales and this scale cannot be simply identified with some obvious physical scale of the problem.  Interestingly, Turing's theoretical insight, which suggested that   ``dappled patterns" might emerge when the reacting chemical species diffuse at different rates, was not actually realized in a laboratory for decades following his proposal. The difficulty was to have the chemical species diffuse at different rates; this was finally achieved in the laboratory in the early 1990's~\citep{castets1990experimental,ouyang1991transition}, when a chemical reactor was made of a gel which only inhibited motion of the (larger) ``activator" molecules.

The Swift--Hohenberg equation~\citep{swift1977hydrodynamic} is a particularly simple  model that demonstrates how patterns form via a symmetry--breaking instability of a spatially uniform state.
The model is a partial differential equation (PDE) for a scalar field $u(x,y,t)$,
\begin{equation}\label{eq:she}
    \frac{\partial u}{\partial t}= r u - (k_0^2 +\nabla^2)^2u +f(u),
\end{equation}
where  $\nabla^2\equiv \frac{\partial^2 }{\partial x^2}+\frac{\partial^2 }{\partial y^2}$ is the Laplacian, and $f(u)$ is a suitable nonlinear function of $u$.
The PDE was originally introduced by Swift and Hohenberg, with the nonlinearity $f(u)=-u^3$, as a simplified model of Rayleigh-B\'ernard convection. In that setting, the scalar field $u$ describes the mid-plane vertical velocity of a thin layer of fluid, heated from below, and the driving from an imposed temperature gradient is captured by the control parameter $r$. The linear stability of $u=0$ to spatially periodic perturbations  $e^{ikx}$ is determined by examining their growth rate $\sigma_k=r-(k_0^2-k^2)^2$, which is  a simple quartic polynomial in $k$ with a maximum at $k_0$. It follows that for $r<0$, the heat conduction state $u=0$ is linearly stable, and for $r>0$, the conduction state is unstable and gives way to a convection pattern. In this case, the  characteristic wavelength of the pattern is $\sim 2\pi/k_0$ near onset, and in the case of Rayleigh-B\'enard convection this scale is set by the thickness of the fluid layer. 

As illustrated in Section~\ref{sec:models:examples}, many models of vegetation pattern formation have features that are qualitatively similar to the Swift--Hohenberg equation (including in some cases the quartic growth functions $\sigma_k$).  These similarities are only expected to be valid in a neighborhood of parameter space near the pattern forming instability, {\it e.g.} near $r=0$ for $|u|\ll 1$. 
Validating models of vegetation pattern formation is challenging when they are intended to go beyond a Swift-Hohenberg model.
Not least of these is that, unlike Rayleigh-B\'enard convection, controlled laboratory experiments do not play a role, due to the long timescales and large spatial scales of community-scale ecosystems. Nonetheless, the power of a generic mechanism, which typically operates for parameters near some stability threshold, resides in its being largely insensitive to the details. This approach is described in section~\ref{sec:models:warning}, where thresholds are often associated  with these special parameter sets near critical transitions.

\subsection{Critical Transitions and the Dream of Early Warning Signs}
\label{sec:models:warning}
\begin{figure}
    \centering
    \includegraphics[width=0.95\linewidth]{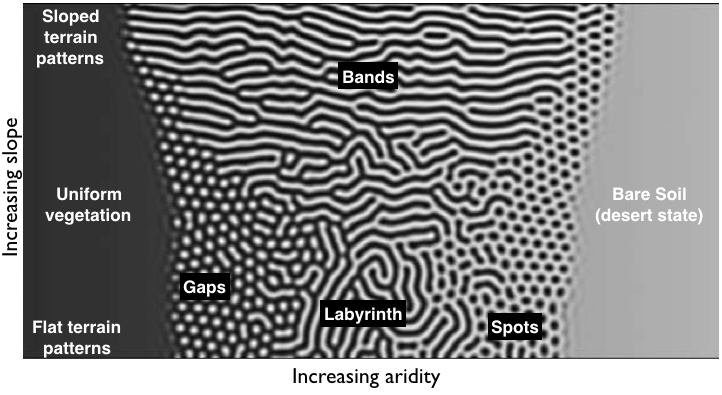}
    \caption{Simulated states in an aridity-slope plane.  Uniform vegetation is associated with high water availability, and bare soil  with low water availability. Annotated figure reproduced from \cite{deblauwe2011environmental}.}
    \label{fig:collapse}
\end{figure}

One framing of the potential significance of dryland vegetation pattern formation is in the setting of  early warning signs of desertification and ecosystem collapse. For instance, in many mathematical models, as the precipitation parameter decreases, there is a well-defined sequence of transitions in the pattern morphology prior to complete collapse to a barren desert state~\citep{deblauwe2011environmental,gowda2016assessing,meron2004,rietkerk2004self,von2001diversity}. This is illustrated in Figure \ref{fig:collapse} in the particular case of the ``flat terrain patterns", where, with increasing aridity uniform vegetation gives way to a matrix of bare patches called ``gaps". With further increases in aridity stress, the gaps merge to a labyrinthine pattern that eventually disintegrates to a new matrix, now of vegetation ``spots", before its ultimate collapse to bare soil.  The pattern sequence can exhibit significant hysteresis,  due to the nonlinear feedbacks that favor bistability of states. Thus it is not easy to reverse a transition from one type of state to another, and in this case, the spot pattern, as precursor to complete collapse, would serve as an ``early warning sign" of  desertification. A perspective piece by 
\cite{rietkerk2004self} identifies other ecosystems, including  peatlands and savannas that may show the same pattern sequence when subjected to increasing environmental stress. 

\begin{figure}
    \centering
    \includegraphics[width=0.9\linewidth]{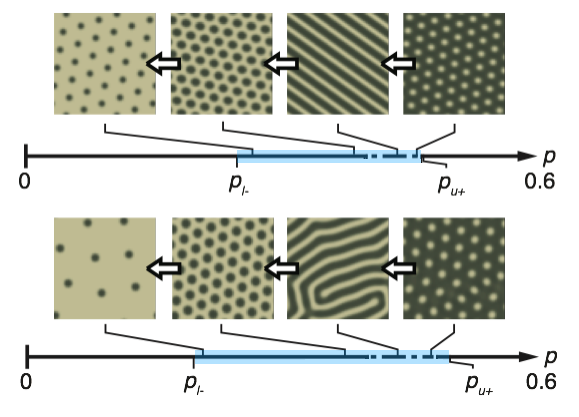}
    \caption{Two examples of the ``standard" flat terrain pattern sequence for changing (dimensionless) precipitation parameter $p$ of the Rietkerk model, modified from Fig. 7 of \cite{gowda2016assessing}; see Section~\ref{sec:models:examples} for description of the Rietkerk model. Light regions are bare soil and dark is vegetation.  The range of precipitation associated with patterns ($p\in[p_{l-},p_{u+}]$), is highlighted in blue, and is larger for the bottom figure than the top due to a change in diffusion ratios. Note that the stripe pattern on the top row distorts to be labyrinthine on the bottom row.}
    \label{fig:karnasprsafig}
\end{figure}

The similarity between gap/spot patterns and patterns of hexagonal Rayleigh-B\'enard convection cells suggests that the same generic mathematical pattern formation framework might be applied to both. In this analogy, convection roll patterns would be supplanted by the labyrinthine patterns. (See Figure~\ref{fig:karnasprsafig}.) Exploiting this analogy, 
\cite{gowda2014transitions} determined the minimal mathematical ingredients for the pattern sequence gaps$\to$labyrinths$\to$spots.  It required a change of sign of a quadratic coefficient in a bifurcation problem, as the precipitation parameter decreased. The ideas were firmly grounded in a small amplitude, weakly nonlinear theory of pattern formation for which retaining only the leading quadratic and cubic terms in a Taylor expansion can capture nonlinear essentials. As such, though, it can make no claims about large amplitude, ``fully nonlinear" patterns. 
In a follow-up paper to this framing, 
\cite{gowda2016assessing} tested it in the setting of the Rietkerk model, described in Section~\ref{sec:models:examples}, by performing both the weakly nonlinear analysis, and extensive numerical simulations outside of the small amplitude regime. In the setting of this particular model, they were able to test the validity of their theoretical quadratic coefficient ``proxy" for the pattern sequence, and  trace the ecological origin of the sign change to key feedbacks incorporated in the model.  Thus, at least for the Rietkerk model, they showed that the gaps$\to$labyrinths$\to$spots sequence  is robust, i.e. insensitive to changes in other (fixed) model parameters. 

While a number of different mathematical models show a gaps$\to$labyrinths$\to$spots sequence as the precipitation parameter is decreased (holding other parameters fixed), it is a challenge to observe this sequence of transitions, at a given location over time, in nature. Perhaps the best observational evidence for this sequence is by 
\cite{deblauwe2011environmental}. In their study they classified vegetation patterns in a 22,255 km$^2$ region that has a strong precipitation gradient, north to south, as it transits out of the Sahel in Sudan. This pattern sequence relies strongly on the terrain being sufficiently flat, as indicated in Figure~\ref{fig:collapse}. In many regions, including the patterned region of Sudan, the flat terrain patterns turn to banded ones transverse to the grade~\citep{deblauwe2011environmental} when there is even a very modest slope. Thus the possibility of a spot pattern as a warning sign of ecosystem vulnerability may vanish with this change in terrain. For possible early warning signs in the context of banded patterns, as precipitation decreases, simulations and analysis of mathematical models have suggested that band spacing may increase~\citep{siteur2014beyond}, the width of vegetation within the bands may decrease~\citep{yizhak2005}, and that the speed of fronts, recovering from disruption, might slow~\citep{zelnik2018regime}.

\begin{figure}
    \centering
    \includegraphics[width=\linewidth]{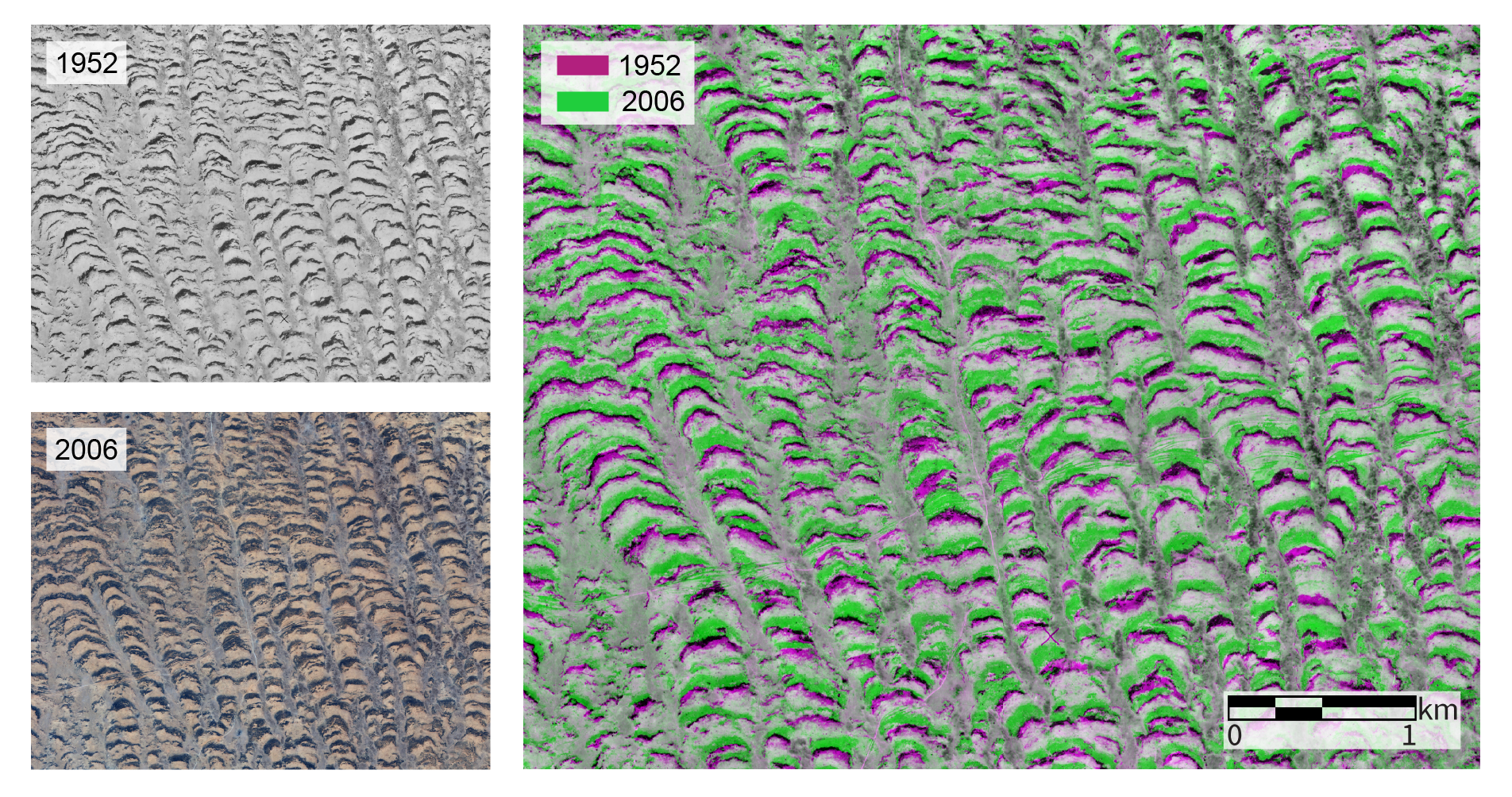}
    \caption{Near Dhahar, Somalia (9.72$^\circ$N, 48.56$^\circ$E). Feb. 1952 image courtesy of the Bodleian Library; March, 2006 image  courtesy of the DigitalGlobe Foundation; both taken in the Jilaal dry season (Dec.-Mar.).
Geo-referencing and false color overlay by Karna Gowda.}
    \label{fig:imagecontest}
\end{figure}

The early warning signs framework relies on identifying the appropriate generic mechanism associated with the critical transitions. Often this exists as an asymptotic (long time) state rather than the transient one associated with, say, a $\sim$decadal timescale. Simulations of mathematical models investigate the response to stress on the system, usually by varying the precipitation parameter with everything else fixed,  and trusting it is not too sensitive to heterogeneity in the environment. One expects, though, that a gradual change in aridity will change the composition of the plant community.  
Interestingly, when one looks at long term change of individual vegetation bands in the Horn of Africa, much is strikingly unchanged, and often the changes that are noted do not align with a clean story from models. For example, a recent study by
\cite{gowda2018signatures} measured the change of thousands of individual vegetation bands, between their 1950's aerial photographs and their modern satellite images. They found that the bands were surprisingly unchanged in regions of Somalia where there were no signs of human impacts. Moreover, in the human impacted areas, the most striking change to the bands that remain intact is that they thickened. Some of this band widening is evident in Figure~\ref{fig:imagecontest}, which is a false color overlay of a 1950's aerial image, geo-referenced with a modern satellite one. The upslope migration of the bands, at a rate predicated by 
\cite{hemming1965vegetation}, is also evident in this overlay. 

We note that there is an alternative approach to early warning signs, not based in pattern, but instead based on a statistical physics analysis of the scaling distribution of vegetation patch size, and its deviation from a power-law distribution~\citep{kefi2007spatial}.
Spatial heterogeneity has been noted in this case to impact the performance of early warning indicators~\citep{genin2018spatially}. For a more comprehensive review of early warning signs in the context of spatial patterns in ecological systems, see~\citep{kefi2014early}.

\subsection{Mathematical mechanisms underlying  pattern formation in vegetation models}
\label{sec:models:mechanisms}

The work of 
\cite{gierer1972theory} provides a conceptual framework for interpreting Turing's mathematical results on diffusion driven instability in terms of short-range facilitation and long-range inhibition.
Many models of spatial pattern formation in ecology, and more broadly, now commonly rely on this as a general principle for  generating pattern forming instabilities in which a uniform state loses stability to perturbations of some finite wavelength~\citep{levin1985pattern}.   This interpretation leads to a generalization of the Turing mechanism described in Section~\ref{sec:models:patternformation}. Some heuristic insight can be gained by considering a small local perturbation of the slowly diffusing activator and rapidly diffusing inhibitor. The local increase in the activator leads to local increase in both the activator and inhibitor.  The inhibitor will spread more quickly  and depress the concentration of both the activator and inhibitor surrounding the local disturbance.  If the diffusion rates are sufficiently disparate, the inhibitor can leave the local perturbation without suppressing it and the perturbation can grow because an excess of the activator remains.  This type of instability is often invoked in vegetation models under an assumption that water ``diffuses" much more quickly than biomass on flat terrain. Biomass, due to cooperative affects of increased infiltration, acts as the ``activator". 
The ``lack-of-water" (as a positive quantity) would  then be considered an inhibitor~\citep{meron2018patterns}.\footnote{For example, a transformation from the water field $W$  to a ``lack-of-water" field $W_L=W_0-W$ in the Klausmeier model~\citep{klausmeier1999regular}, for some constant $W_0$ such that $W_L>0$, can lead to an activator-inhibitor system for $(B,W_L)$, where $B$ is the biomass field.}

A generalization of the Turing instability can appear in kernel-based models where the temporal dynamics at a location are governed by integral terms that capture influence from the neighborhood surrounding a particular location.  For example, this is often used to capture plant growth for cases when root systems harvest water over an extended area surrounding a given plant. This provides long range inhibition while short range facilitation can be obtained through moisture trapping over smaller areas. From the perspective of symmetry-breaking, the resulting bifurcations  are indistinguishable from Turing bifurcations. The dynamics of both types of systems can therefore be described by the same class of reduced models in appropriate limiting cases.

A variation of this general framework for pattern formation is further exemplified by models of banded vegetation patterns on hillslopes.  In this case the water undergoes fast advection instead (i.e. downslope flow) of fast diffusion, leading to a so-called Turing-Hopf instability~\citep{rovinsky1992chemical}.  An immediate consequence of spatial asymmetry, between uphill and downhill, is that patterns are expected to travel and might then be captured in models as  traveling wave patterns.
Vegetation models such as Klausmeier's~\citep{klausmeier1999regular}  predict the travel direction to be uphill when advection is introduced, consistent with some of the observational evidence at a number of locations \citep{deblauwe2012determinants}.
In a Turing system where a pattern is generated by diffusion, an asymmetry can also be introduced, e.g. by adding an advective term.  Asymmetry can also be incorporated into kernel-based non-local models, typically by introducing asymmetry into the integration kernel.  Both of these cases generate similar behavior of traveling wave patterns that can model banded vegetation with an upslope colonization.   

The perspective discussed in this section focuses on the ``formation" of the pattern at a particular point in parameter space through an instability of a uniform state.  There are a number of possible scenarios by which an ecosystem can evolve in time to a particular patterned state.   Environmental changes could bring a once stable uniform state to an unstable region of parameter space and the pattern could grow from perturbations or fluctuations.  For instance, a decrease  in water availability could lead to the emergence of patterns from small-scale variation in an initially uniform  vegetation state.  Another possibility is that the vegetation patterns colonize a region of bare soil via an invasion front. See~\cite{samuelson2018advection} for a bare bones  model that illustrates the latter scenario.  Studies of front dynamics in vegetation models have suggested restrictions on what scenarios are possible~\citep{zelnik2017desertification}. The prediction in this case is that the colonization of bare soil by vegetation patterns may not be gradually reversed by a receding front, and the patterns are instead lost all at once through a uniform collapse to bare soil. Other studies have suggested that the pattern selected as a result of colonization may be more consistent with empirical data than the patterns formed from small perturbations of a uniform state~\citep{sherratt2015using}.

\subsection{Development of  Mathematical Models for Vegetation Pattern Formation}
\label{sec:models:development}

The first models of vegetation pattern formation seem to appear around the mid 1990's~\citep{mauchamp1994simulating,thiery1995model}, and are discrete-time and discrete-space cellular automata models.   The state of each location on a spatial grid, at each time, is determined, via some set of defined rules, by the state of itself and its neighbors at the previous time.  The state often captures characteristics about both biomass and water at a given point in space and time.  The models can be extremely mechanistic in the way they capture the various processes involved, leading to a large number of parameters that must be determined from data.  This modeling approach tends, therefore, to be very closely connected to observation and often includes comparison to field studies.  

The first efforts to incorporate ideas from pattern formation come from
\cite{lefever1997origin}.  
On a historical note, Lefever made important contributions to the study of pattern formation in nonlinear optics~\citep{lugiato1987spatial} and in chemical systems. Together with Prigogine, he formulated the Brusselator model~\citep{prigogine1968symmetry}.
The Lefever and Lejeune model is a conceptual model that captures dynamics of a continuous biomass field using a integro-differential equation accounting for plant-to-plant interactions mediated by the environment.  Other work from this branch of modeling has focused on spatially localized patches of vegetation as examples of dissipative structures~\citep{lejeune2002localized}.

Explicit modeling of environmental dynamics by coupling a continuous biomass field to a continuous water field in a pair of reaction-advection-diffusion equations is first introduced by 
\cite{klausmeier1999regular} and later analyzed in detail by 
\cite{sherratt2005analysis,sherratt2007nonlinear}.  Klausmeier's original model is formulated for banded vegetation patterns on a uniform hillslope.  Recent work has generalized the model so that it can capture flat-terrain patterns as well~\citep{van2013rise}. Using one model to generate both flat-terrain and uniform hillslope patterns, a goal first introduced by
\cite{von2001diversity}, may sometimes be appropriate.  An interesting case where they appear in proximity is in the region of Sudan studied in \citep{deblauwe2011environmental}.  It is not the case that the two types of patterns always occur in association, e.g. flat terrain patterns have not been described in the Horn of Africa, where banded patterns are prevalent~\citep{gowda2018signatures}.  It is an open question whether the mechanisms are fundamentally different between formation of flat-terrain and of banded hillslope patterns.

Contemporary reaction-advection-diffusion models are moving towards capturing  more mechanistic details of various processes thought to play a key role in pattern formation.  For example,  the model due to 
\cite{hillerislambers2001vegetation,rietkerk2002self}, which we describe below as "the Rietkerk model," subdivides the water into a surface water and soil moisture to better capture the biomass feedback on the infiltration of water from the surface into the soil.  
\cite{gilad2004ecosystem} also incorporate this three-field (two water and one biomass) modeling approach.  Nonlocal seed dispersal has been explored~\citep{bennett2018long,pueyo2008dispersal,thompson2008role}.
Some focus on the effects of seasonality or stochasticity of rainfall~\citep{kletter2009patterned,konings2011drought,siteur2014will}.  Others couple the hydrological processes to landform evolution via sediment transport dynamics~\citep{saco2007eco}.  There is also now a push to model the interaction of multiple species of vegetation within these patterned ecosystems~\citep{gilad2007dynamics}.  The influence of grazers~\citep{van2002spatial,siero2018nonlocal} and interactions with termites~\citep{pringle2017spatial} have been considered as well. Even hybrid models that capture plants with an agent-based approach and couple them via a continuous water field have been proposed as a way to incorporate individual plant characteristics~\citep{vincenot2016spatial}.

\subsection{Examples}
\label{sec:models:examples}

We now describe three examples from the literature that serve as foundations for many contemporary reaction-advection-diffusion models of vegetation patterns. We assume a two-dimensional spatial domain throughout this discussion, though many model studies restrict to one spatial dimension for simplicity. The Lefever--Lejeune model~\citep{lefever1997origin}, perhaps the first to take a pattern formation perspective, captures self-interactions of biomass through nonlocal integral kernels that exhibit short-range facilitation and long-range inhibition.   The 
\cite{klausmeier1999regular} model explicitly models water dynamics coupled to biomass evolution on a uniform hillslope in a pair of equations reminiscent of the Gray--Scott chemical reaction model~\citep{gray1983autocatalytic}. Pattern formation in this case occurs as a result of a fast downhill advection of water and slow diffusion of biomass, and does not produce patterns in absence of a hillslope.  A model by Rietkerk and collaborators~\citep{hillerislambers2001vegetation} subdivides water into a fast-moving surface flow and a slowly-diffusing soil moisture that are coupled to a slowly-diffusing biomass field in order to better capture biomass feedback on infiltration of water from the surface into the soil.
The increased complexity of the Rietkerk model relative to the previous two examples illustrates the trade-offs between fidelity and parsimony in models of this type. See \cite{zelnik2013regime} for a more detailed mathematical comparison of versions of these models to a version of a model by 
\cite{gilad2007mathematical}. 

\subsubsection*{Lefever--Lejeune Model}
The Lefever--Lejeune model~\citep{lefever1997origin} has a very simple structure, with the dynamics of biomass field $B(\Xbf,T)$ depending on a growth term $\FF_G$ multiplied by an interaction term $\FF_I$ minus a mortality term $\FF_M$. 
\begin{equation}\label{eq:LL}
   \partial_T B=\FF_G \FF_I-\FF_M
\end{equation}
Each of these terms represents an integral that captures nonlocal interaction between the plants:
\begin{align*}
    \FF_G(\Xbf,T) &=\int  \lambda w_G(\Sbf) B(\Xbf+\Sbf,T) \left(1+\Omega B(\Xbf+\Sbf,T)\right) d\Sbf \\
    \FF_I(\Xbf,T) &= 1 - \int  w_I(\Sbf) \frac{B(\Xbf+\Sbf,T)}{K}  d\Sbf\\
    \FF_M(\Xbf,T) &=\int  \eta w_M(\Sbf) B(\Xbf+\Sbf,T)  d\Sbf.
\end{align*}
Here the integrals are over a two-dimensional spatial domain, i.e. both $\Xbf$ and $\Sbf$ represent two-dimensional vectors.
The ``replication constant" $\lambda$ captures the growth rate of the plants, the ``cooperativity constant" $\Omega$ characterizes plant facilitation, the ``biomass packing limit"  $K$ plays the role of a carrying capacity,  and the ``mortality constant" $\eta$ characterizes biomass loss.  The integration kernels $w_\alpha$ with $\alpha=G,I,M$ are typically assumed Gaussian, and on a two-dimensional spatial domain, take the form
\begin{equation*}
    w_\alpha(\Sbf) = \frac{1}{2\pi L_\alpha^2}e^{-|\Sbf|^2/2L_\alpha^2}
\end{equation*}
The lengthscales responsible for the pattern-forming instability are contained in a growth kernel $w_G$ in the growth term $\FF_G$, with a characteristic ``dissemination lengthscale" $L_G$ for seed dispersal, and a competition kernel $w_I$ in $\FF_I$, with a characteristic ``inhibition length" $L_I$.  There is also a ``toxicity lengthscale" $L_M$ associated with the mortality kernel $w_M$, but it is not expected to play a role in the mechanism for pattern formation, and is often taken to be zero (i.e. $\FF_M=\eta B$).

With $L_M=0$ and in the limit of low biomass densities and small biomass gradients\footnote{The key role of these assumptions is to ensure that substituting a Taylor expansion of the biomass into the integrals  provides good approximations.}, the equation can be approximated by the fourth-order PDE
\begin{equation}
    \partial_t b =(1- \mu)b +(\Lambda-1)b^2 - b^3 + \frac{1}{2}(L^2-b)\nabla^2 b -\frac{1}{8}b\nabla^4 b
\end{equation}
where time has been rescaled by birth rate $t=\lambda T$, space has been rescaled by inhibition length $\mathbf{x}=\Xbf/L_I$ and biomass has been rescaled by carrying capacity $b=B/K$.  The dimensionless parameter $\Lambda=K\Omega$ measures availability of resources and $\mu=\lambda/\eta$ is a birth-to-death ratio. This reduced equation shares many qualitative characteristics with the Swift--Hohenberg equation~\eqref{eq:she}; indeed a variant of the model~\eqref{eq:LL} reduces exactly to an equation of Swift--Hohenberg type in an appropriate limit~\citep{lefever2009deeply}.

The Lefever--Lejeune model does not explicitly track the dynamic evolution of water.  It does, through the choice of parameters and functional form of the kernels, implicitly make assumptions about the environment.  Because the model does not attempt to capture details of hydrology or environment, the assumptions about these are minimal.  On the other hand, it can be difficult to use this model to study  the influence of the environment since it is often unclear how any given  process is encoded.  For example, rainfall is not explicitly included as a parameter, but $\mu$ indirectly captures the influence of aridity under the assumption that plants lose biomass at a higher rate in more arid environments.  However one might expect that $\Lambda$, which characterizes resource availability, is effected by aridity as well.  As another example, the effect of a uniform hillslope can be captured in the model by introducing anisotropy in a weighting function that breaks symmetry along the uphill-downhill direction.  

\subsubsection*{Klausmeier Model}

The Klausmeier model~\citep{klausmeier1999regular} couples the dynamics of the biomass field, $B$, to the dynamics of a water field, $W$, on a uniform hillslope where uphill is in the $+X$ direction:
\begin{align}
       \partial_T W &= \underbrace{V\partial_X W}_{\text{transport}} - \underbrace{L W}_{\text{evaporation}}   - \underbrace{R W B^2}_{\text{transpiration}} + \underbrace{P}_{\text{precipitation}} \label{eq:klaus:w}\\
        \partial_T B  &= \underbrace{D \nabla^2 B}_{\text{dispersal}} - \underbrace{M B}_{\text{mortality}}  + \underbrace{JR WB^2}_{\text{growth}} \label{eq:klaus:b}
\end{align}
Water is input into the system with precipitation rate $P$, assumed to be the constant annual mean. The advection rate, $V$, depends on the slope of terrain and represents the  overland water flow; $V>0$ means the hill slopes upwards in the $+X$-direction and the water is transported in the opposite downslope direction.  The soil evaporation rate is given by $L$.  The transpiration, $RWB^2$, drives biomass growth with a water-use efficiency parameter $J$.  The quadratic dependence on biomass in the growth captures, heuristically, the positive biomass feedback on infiltration of water into the soil.  The mortality rate $M$ captures biomass loss due to both natural death, but also due to grazing and to human pressure.  Seed dispersal is modeled through linear diffusion with rate $D$. 

\begin{figure}
    \centering
    \includegraphics[width=\textwidth]{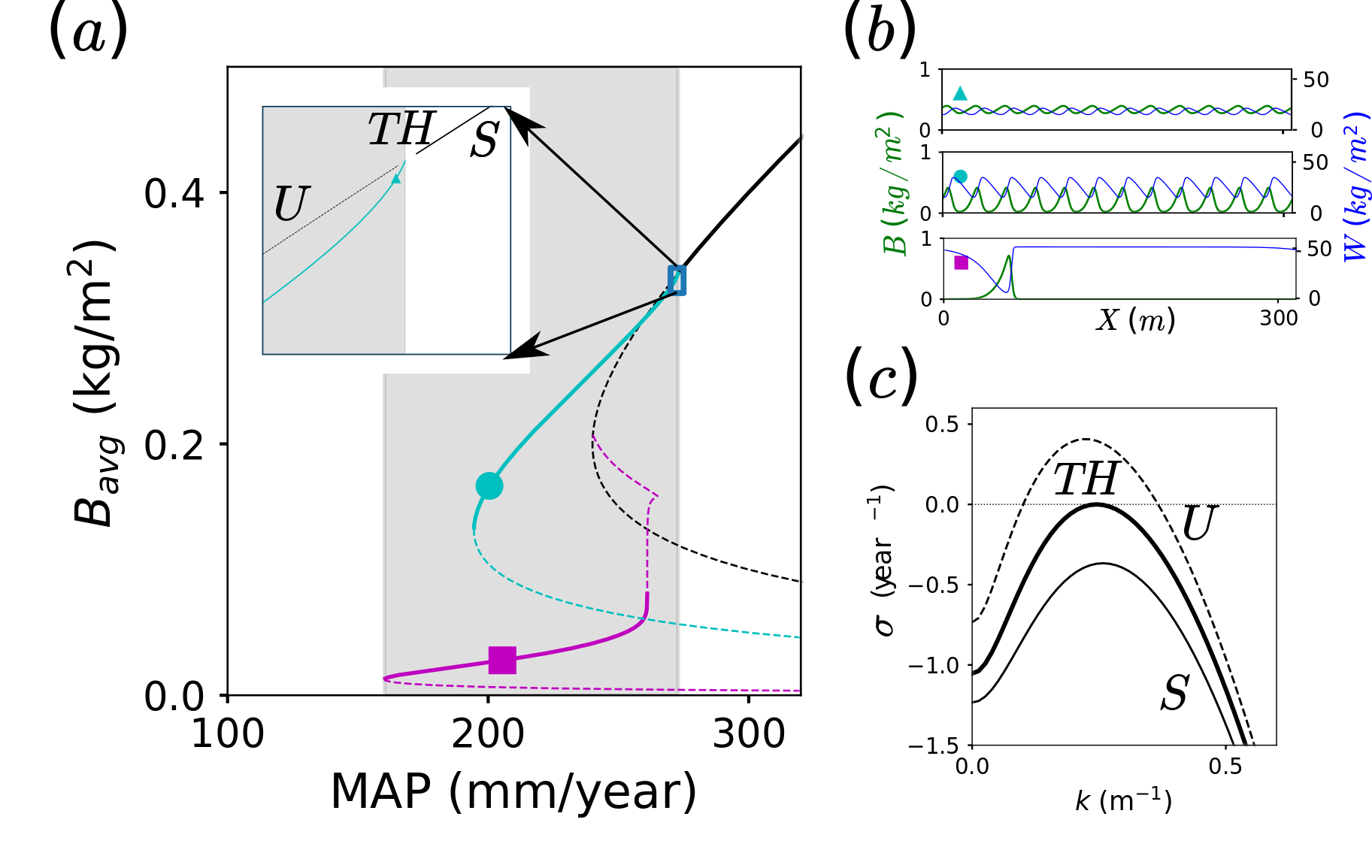}
    \caption{(a) Bifurcation diagram showing average biomass as a function of mean annual precipitation (MAP) for the uniform vegetation state (black), the Turing-Hopf pattern (cyan) and the one-pulse ``oasis state" (magenta).  The inset shows a zoom in near the Turing-Hopf bifurcation. The branches are solid where the solutions are linearly stable and dashed where the solutions are linearly unstable.  (b) solutions profiles for biomass (green)  and water (blue) are shown at points indicated in the bifurcation diagram. 
    (c) Growth rate $\sigma$ of perturbations $e^{ikx}$ to uniform vegetation states; here linearized about stable (solid) and unstable (dashed) uniform states.  The bold line is the linearization at the Turing-Hopf bifurcation where the uniform state is marginally stable.      Parameters of \eqref{eq:klaus:b}: $L=4$ year$^{-1}$,  $M=1.8$  year$^{-1}$,  $J=0.003$ kg Dry Mass / kg H$_2$O, $R=100$ kg H$_2$O/m$^2$/year/ kg$^2$ Dry Mass, $D=10$ m$^2$/year and $V=63$ m/year. 
    }
    \label{fig:bif}
\end{figure}

Figure~\ref{fig:bif} shows a typical bifurcation diagram associated with a pattern-forming vegetation model such as~\eqref{eq:klaus:b}, with shading to indicate the range of mean annual precipitation parameter $P$ where patterns exist. Two important examples of patterned states are shown; one bifurcates at the Turing-Hopf point that determines the right boundary, and the other is a single ``pulse" or ``oasis" state that marks the left boundary of existence. Also, shown in this figure are examples of growth rate functions $\sigma_k$ associated with small periodic perturbations $e^{ikx}$ of uniform solutions.

One advantage of Klausmeier's model over the Lefever--Lejeune equation is that each term in the $W$ equation can be associated with a specific ecohydrological process.  For example, the  topography influences water transport and so it is clear which term is capturing the hillslope's effect. However, because the $W$ field captures a conglomeration of hydrological processes, across  timescales, and both on the surface and in the soil, it is not consistent with the observed hydrology in many ways. For example, the model predicts that the water field peaks in bare soil regions as can be seen in Figure~\ref{fig:bif}(c), yet soil moisture is observed to be higher in the vegetated zones after rain events, as described in Section~\ref{sec:earlyinvestigations}.

The reaction terms of the Klausmeier model are a reparametrization of the Gray--Scott model, a classic activator-substrate system.  The generalized Klausmeier/Gray--Scott model~\citep{van2013rise}, which additionally includes a water diffusion term, can be transformed into the standard form of Gray--Scott model studied by 
~\cite{pearson1993complex} in the case that the water transport is purely diffusive.\footnote{However, the parameters used in the context of vegetation patterns correspond to unphysical parameter choices in the original chemical kinetics setting.}  This purely diffusive version of the model has also been shown to reduce to an equation of Swift--Hohenberg type, i.e. Equation~\eqref{eq:she}, for the pattern amplitude in an appropriate limit~\citep{gandhi2018spatially}.

\subsubsection*{Rietkerk Model}

The model proposed by
\cite{rietkerk2002self,hillerislambers2001vegetation} takes an additional step to remedy the poor hydrological predictions of the Klausmeier model by subdividing the water into a surface water field and soil moisture field.  An ordinary differential equation precursor to this model, aimed at studying the effects of grazing in water-limited and nutrient-limited ecosystems~\citep{van1997catastrophic},  was published in the same year as  Lefever and Lejeune's model. The Rietkerk model for vegetation patterns is given by
\begin{align}
    \partial_T H &=\underbrace{D_h\nabla^2 H}_{\text{surface transport}} -\underbrace{\mathcal{I}}_{\text{infiltration}} +\underbrace{P}_{\text{precipitation}} \\
    \partial_T W &= \underbrace{D_w \nabla^2 W}_{\text{soil diffusion}} -\underbrace{r W}_{\text{evaporation/leakage}} + \underbrace{\mathcal{I}}_{\text{infiltration}} - \underbrace{\mathcal{G}}_{\text{uptake}} \\
    \partial_T B &=\underbrace{D_b\nabla^2B}_{\text{seed dispersal}} - \underbrace{m B}_{\text{mortality}}+ \underbrace{c\mathcal{G}}_{\text{growth}}
\end{align}
where water is input to the surface by precipitation with constant rate $P$ and quickly diffuses along the surface with rate $D_h$. After infiltrating into the soil, water diffuses with rate $D_w$, is lost by evaporation or leakage deeper into the soil at rate $r$, and is lost by plant uptake characterized by the nonlinear function $\mathcal{G}$.  Seed dispersal is modeled by linear diffusion with rate $D_b$, plant mortality from grazing or otherwise occurs at a rate $m$ and growth is given by $c\mathcal{G}$.  
The nonlinearities of the model arise through the growth and infiltration functions, given by
\[  \mathcal{G} = g B\frac{W}{W+k}, \qquad 
 \mathcal{I} = \alpha H\frac{B+ qf}{B+q}. \]
The parameter $g$ sets the maximum growth rate and the parameter $k$ sets the soil moisture required for half that growth rate to occur. The maximum infiltration rate is set by $\alpha$ and the dimensionless feedback parameter $f\in[0,1]$ controls the ratio of the bare soil to vegetated soil infiltration, e.g. it represents the infiltration reduction introduced by a soil crust. Specifically, bare soil infiltrates at reduced rate $\alpha f$, while when the biomass  $B\gg q$, the infiltration approaches its maximum rate $\alpha$.

Separating surface water $H$ from soil moisture $W$ allows for more detailed modeling of the biomass feedback on infiltration of water from the surface into the soil, a process long thought to play an important role in pattern formation of these ecosystems, as described in Section~\ref{sec:earlyinvestigations}.
While this improves predictions regarding soil moisture distribution, the surface water is still problematic to interpret.  In order to account for the assumption that water is input at a constant rate year-round, the infiltration rate is reduced by an extra factor of the fraction of time it rains during the year.   
Even with this caveat, the Rietkerk model provides a sound framework for exploring predictions from extensions that capture additional processes in more detail.  However, more detailed modeling comes at the cost of increased complexity. The scaled version of the Rietkerk model has six dimensionless parameters, as opposed to three for the dimensionless versions of the Klausmeier and Lefever--Lejeune models.

\subsection{Challenge: Constraining Parameters in Parsimonious Models}
\label{sec:models:parameters} 
Models appear along a spectrum in terms of level of details about the underlying processes, but can be broadly classified as either ``conceptual" or ``mechanistic."    Mechanistic models attempt to capture the actual processes with high fidelity, but this often leads to increases in mathematical complexity and number of parameters. Despite the parsimony associated with conceptual models, they are able to reproduce the observed patterns and are often amenable to mathematical  analysis.  However, the connection between the model parameters and the actual processes being modeled can be tenuous. Constraining model parameters based on data can therefore be challenging and the values obtained can be difficult to interpret.  Moreover, the parameter values associated with a given process can vary over orders of magnitude between models.      
The typical approach to estimating parameters in the more conceptual models involves a combination of constraints based directly on observation or measurement, and inferred constraints that rely heavily on model assumptions.  The inferred constraints can be based on characteristics of the  bare soil state, uniform vegetation state or patterned state.  

We now illustrate the process for choosing parameter values within the most parsimonious of models within the reaction-advection-diffusion framework, following the approach of~\cite{klausmeier1999regular}. A similar approach based on observed pattern characteristics and linear predictions has been used to constrain two key dimensionless quantities in the Lefever--Lejeune model~\citep{lejeune1999short}.  More detailed studies have also attempted to fit the plant interaction kernel using data from measurements of the root system~\citep{barbier2008spatial}. 

\begin{figure}
    \centering
    \includegraphics[width=0.4\textwidth]{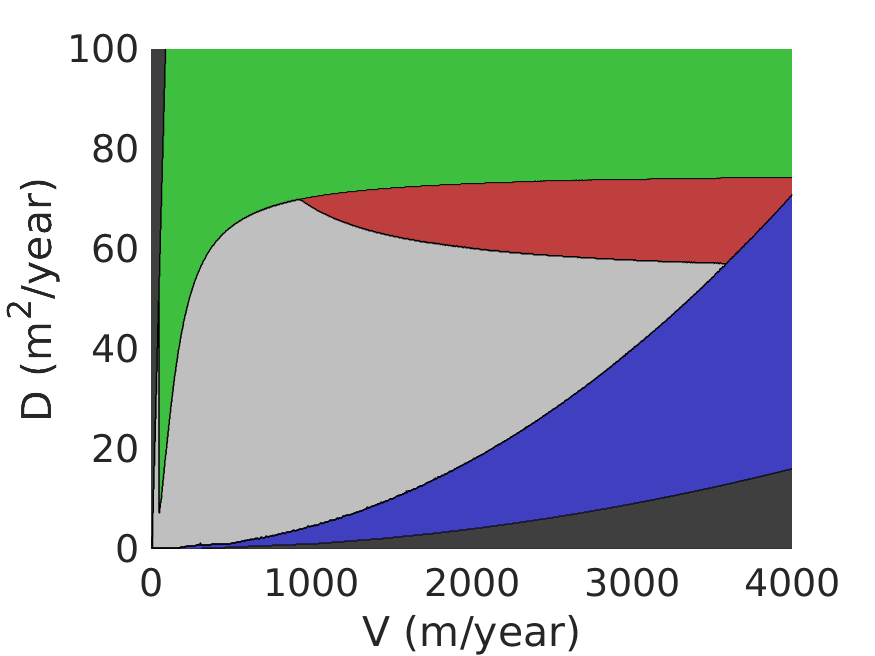}\hspace{1cm}
    \includegraphics[width=0.4\textwidth]{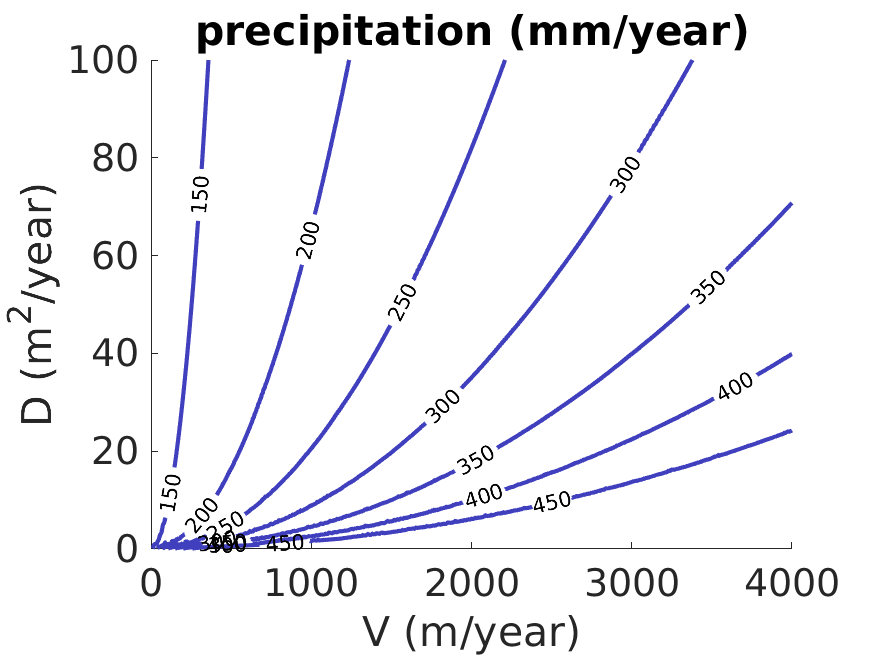}\\ 
    \includegraphics[width=0.4\textwidth]{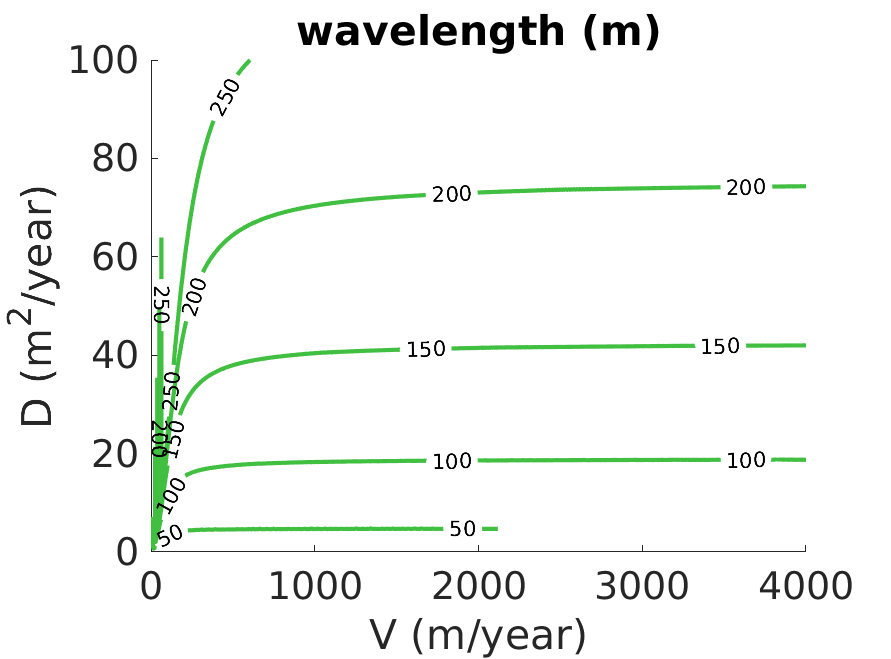}\hspace{1cm}
    \includegraphics[width=0.4\textwidth]{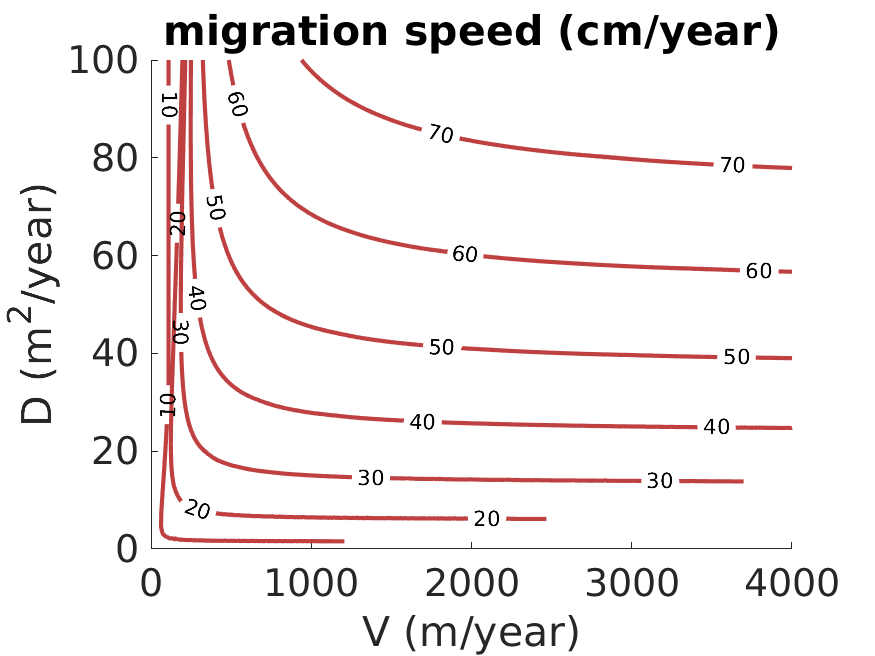}    
    \caption{Constraining parameters in the Klausmeier model. The light gray region in the upper left panel shows a range of parameter values in the $(D,V)$-plane that satisfies the following constraints: $120<$precipitation$<350$ (mm/yr) with blue outside of this range, $10<$wavelength$<200$ (m) with green outside of this range of what remains, and $1<$ migration speed $< 60$ (cm/yr) with red outside of this range of what remains.}
    \label{fig:constrain}
\end{figure}

In  
\cite{klausmeier1999regular}, the constant rainfall parameter $A$ was set to 300 mm/year,  a typical mean annual precipitation for semiarid regions. Values for plant water use $J=0.002$ kg Dry Mass/kg H$_2$O for trees ($J=0.003$ kg Dry Mass/kg H$_2$O for grasses)  and mortality $M=0.18$ year$^{-1}$ for trees ($M=1.8$ year$^{-1}$ for grasses)  were taken from estimates in a previous modeling study~\citep{mauchamp1994simulating}, which were based on data about maximum water use efficiency and carbon costs required to maintain leaves.  
The evaporation rate $L=4$ yr$^{-1}$ is set by typical values of water  from observation and predictions about the bare soil state. The transpiration rate $R=1.5$ mm H$_2$O/yr/(kg dry mass)$^2$ for trees ($R=100$ mm H$_2$O/yr/(kg Dry Mass)$^2$ for grasses) is set by typical biomass values from observation and predictions about the uniform vegetation state.
This leaves the two transport parameters to estimate. One way to do this would be to constrain the water advection parameter $V$ and the biomass diffusion parameter $D$ so that the the wavelength $\lambda$ and migration speed $c$ of the patterns agree with  typical observed values. Based on observations by~\cite{deblauwe2012determinants}, we use $\lambda=10-200$ m and $c<60$ cm/yr and assume onset of patterns occurs with MAP of 120-350 mm/yr. Figure~\ref{fig:constrain} shows that this constrains both parameter to approximate ranges of $V\sim 5-3500$  m/year and $D\sim 1-70$ m$^2$/year for trees.  The ranges for grasses are more constrained: $D\sim 1-15$ m$^2$/year and $V\sim 5-15$ m/year. 
For comparison,
\cite{klausmeier1999regular} took $D=1$ m$^2$/year and $V=365$ m/year and typical values for the analogous transport parameters used in the Rietkerk model
are  surface water flow speed of $V=3700$ m/year and biomass diffusion rate of $D_b=37$ m$^2$/year.  We note that typical water flow speeds are actually on the order of meters per second (i.e. $10^7-10^8$ m/year), so these values, based on a model with constant precipitation or ``drizzle", do not align with those following a rain event that might lead to the observed sheet flow  as was described in Section~\ref{sec:earlyinvestigations}.  

The above estimates of the transport parameters rely on predictions about wavelengths and migration speeds of spatially periodic vegetation patterns at the onset of small amplitude patterns, i.e. at the Turing-Hopf bifurcation point where the spatially uniform state loses stability. However, the observed patterns correspond to large amplitude states in the model and appear in parameter regimes where a range of wavelengths are stable.  This predicted multistability can be summarized by the so-called ``Busse balloon", originally developed in the context of Rayleigh-B\'enard convection~\citep{busse1978nonlinear}. These balloons, for example in the $(P,k)$ parameter plane, are based on nonlinear calculations that characterize the boundaries of stability for patterns as a function of the precipitation level $A$ and the pattern wavelength ($2\pi/k$). Such calculations, which are a nontrivial undertaking, have been carried out for  the generalized Klausmeier/Gray--Scott model~\citep{van2013rise}.
Recent progress has been made by generating observational Busse balloons to compare with predictions from this model~\citep{bastiaansen2018multistability}.  For a given site in Somalia (i.e. with fixed aridity) the empirical ``Busse balloons" of  band migration speeds, and of biomass level, are estimated as a function of the local wavelength of the pattern and the local elevation gradient. This allows for comparison between qualitative trends in the data and predictions from a model.  For the generalized Klausmeier/Gray--Scott model, the observations are consistent with predicted trends in biomass and migration speed as a function of the pattern wavelength  but do not match predictions about trends with local elevation grade. 

\section{Opportunities for Modeling: Time-scale Separation and Heterogeneity} \label{timescales}

Section~\ref{sec:models:parameters} highlighted some of the challenges associated with modeling surface water flow using annually averaged quantities.  
Modeling these processes on the appropriate fast hydrological timescale means that model parameters correspond more closely to measurable quantities. 
Capturing fast processes presents a different set of challenges in terms of model complexity and numerical computation time. For example, a study that uses a process based ecohydrological model can make detailed predictions about water dynamics on short timescales~\citep{paschalis2016matching}. However, the simulations are limited by computational complexity to a few decades using static vegetation patterns on an idealized hillslope.  
Section~\ref{sec:scales} summarizes some mathematical techniques that exploit separation of scales in models to simplify analysis and/or generate reduced descriptions. Section~\ref{sec:stochastic} describes the important challenge of capturing  rainfall as a stochastic driver in models of dryland vegetation patterns.  Section~\ref{sec:topography} describes ways that spatial heterogeneity, due to large-scale topographic variations, might provide an opportunity for better connecting observations with models.    

\subsection{Scales in Time and Space} \label{sec:scales}

Many of the reaction-advection-diffusion models described in Section~\ref{sec:models} are formulated on the timescale of biomass growth and death and capture emergent dynamics on a century scale. 
In this section we highlight progress and potential for capturing processes on appropriate scales within multiscale models and then reducing them using formal mathematical techniques.  The advantage, when such reduction is possible, is that it provides a framework for capturing the (sometimes unintuitive) interactions between the fast time and short lengthscale processess on the long time and large scale dynamics.  

Because water and biomass evolve on different temporal and spatial scales in dryland ecosystems, mathematical methods such as ``geometric singular perturbation theory" that rely on separation of scales have been applied in conceptual models by separating the fast dynamics associated with water transport from slow dynamics of biomass dispersal~\citep{bastiaansen2018dynamics,carter2018traveling,sewalt2017spatially,siero2015striped}.   In contrast to Turing-type analysis 
that can identify existence and stability of small amplitude (weakly nonlinear) patterns near onset of an instability, these methods apply for large amplitude  (strongly nonlinear) patterns far from  a pattern forming instability.   The approach is to separately compute the fast (or small-scale) dynamics and the slow (or large-scale) dynamics in reduced systems and then connect these solutions together.  The key to developing reduced descriptions in this approach rests on the result that the solution constructed in the limit that the scales do not interact persists when the interaction between scales is small but nonzero.

Multiple scales methods are already used for dealing with the timescale separation between water transport and biomass dispersal which are transport terms in the model.  Such methods could also be used to capture the timescale separation that occurs within the interaction terms, between the slow biomass growth and death rates and the fast water infiltration rate.
In the context of geometric singular perturbation theory, this would suggest using a pair of reduced models: one for during and shortly after rain events in which biomass is fixed and the water fields evolve on a fast time, and one for after the surface water completely infiltrates into the soil in which soil moisture and biomass evolve on a slow timescale. The complete dynamics could then be approximated by ``gluing" the solutions from these two systems together.

\subsection{Stochastic Drivers}\label{sec:stochastic}

\begin{figure}
\begin{center}
\includegraphics[trim={2cm 2cm 2cm 0cm},width=0.8\linewidth]{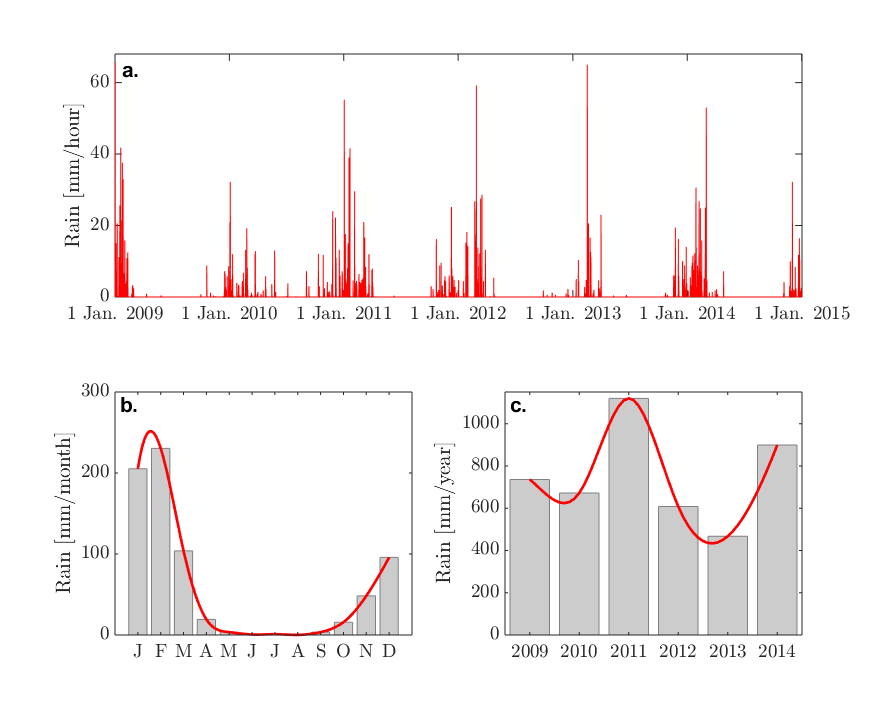}
\end{center}
\caption{Example of daily, seasonal, and interannual rainfall variability: (a) temporal sequence of rainfall events (hourly data from January 2009 to December 2014), (b) mean monthly rainfall, and (c) total annual rainfall at Sturt Plains, Australia (data from  \cite{beringer2016introduction}). }
\label{fig:rainfall}
\end{figure}

\cite{NoyMeir1973} defined arid and semiarid ecosystems as ``water-controlled ecosystems with infrequent, discrete, and largely unpredictable water inputs". Indeed, the timing and magnitude of rainfall events in these areas has a large random component \citep{Nicholson1980,NoyMeir1973,Rodriguez2007}.  While energy inputs (e.g., radiation and temperature) exhibit a fairly continuous variation over the year, rainfall usually comes in discrete ``pulses" of very short duration \citep{Huxman2004,NoyMeir1973,Rodriguez2007,Schwinning2004,schwinning2004thresholds} (see Figure \ref{fig:rainfall}.)
The intermittent and unpredictable character of rainfall results in fluctuations in soil moisture availability, which controls vegetation and ecosystem functioning (e.g., resistance to drought, productivity, germination). The response of the system  depends on the magnitude of each pulse as well as the time interval between rainfall events, namely rainfall depth and frequency \citep{Rodriguez2007}.  

The role of stochastic rainfall variation in vegetation pattern formation is still poorly understood, thus opening up a number of possibilities for the development of modeling frameworks that explicitly account for the role of environmental stochasticity~\citep{Borgogno2009,Rohani1997}. 
Moreover, predicted changes in rainfall frequency and intensity under climate change \citep{ipcc2013,Pascale2016,Seager2007,Solomon2009} may  dramatically impact dryland ecosystems. Understanding how these environmental fluctuations impact vegetation self-organization will help us assess the potential impact of a changing climate
on these vulnerable ecosystems, and their possible desertification~ \citep{rietkerk2002self,kefi2007spatial,paschalis2016matching}.

In some cases, vegetation models of the type described in Section~\ref{sec:models} have been modified to include random environmental drivers~\citep{kletter2009patterned,siteur2014will} or augmented by coupling to  climactic conditions~\citep{konings2011drought}  as a way of studying the influence of stochastic rain ittermitancy and drought on deterministically formed patterns.

Noise-induced pattern formation (see 
\cite{sagues2007spatiotemporal} for a review from a physics perspective) has also been posited alternate mechanism for vegetation patterning.
\cite{Dodorico2006} show that the random switch between wet and dry conditions associated intermittent rainfall may trigger pattern formation even in the absence of a pattern forming bifurcation. In fire-prone savannas, the emergence of vegetation patterns is explained by purely noise-induced mechanisms linked to fire randomness and fire-vegetation feedbacks \citep{Dodorico2007}.

\subsection{Topographic Heterogeneity}
\label{sec:topography}

Environmental heterogeneity  occurs on many different scales and its influence can often be observed through characteristics of the vegetation patterns. 
For instance, topographic variation on the scale of an individual plant, microtopography, has been investigated in connection to a transition from banded vegetation patterns to vegetation following irregular drainage patterns by 
\cite{mcgrath2012microtopography}. Their model predicts that microtopography can disrupt the formation of vegetation bands and lead to an irregular drainage-type pattern of vegetation aligned along the slope.   Other model studies have additionally sought to capture the influence of vegetation patterns in the formation of terracing of the underlying terrain along the hillslope through a coupling with erosion and soil transport dynamics~\citep{baartman2018effect,saco2007eco}. 
 
The pattern formation models of Section~\ref{sec:models} are posed for ecosystems that evolve over time and space deterministically. However, because the timescales for the biomass birth and death processes are so long, it is challenging to test the model predictions through observations via time-lapsed images. An alternative to testing predictions against time-series is provided by the natural environment with its topographic heterogeneity. 
As described in Section~\ref{sec:observations}, topography has long been known to play an important role for vegetation patterns. 

Topographic data at 30-90m resolution~\citep{Farr:2007ib} is  available worldwide, which is a comparable spatial scale to some of the individual vegetation bands. Can we exploit this natural environmental driver,  with all its varied realizations over space, to test and improve the models? Arcing of vegetation bands, described below, provides an example of how this might proceed. 

\begin{figure}
    \centering
    \includegraphics[width=\textwidth]{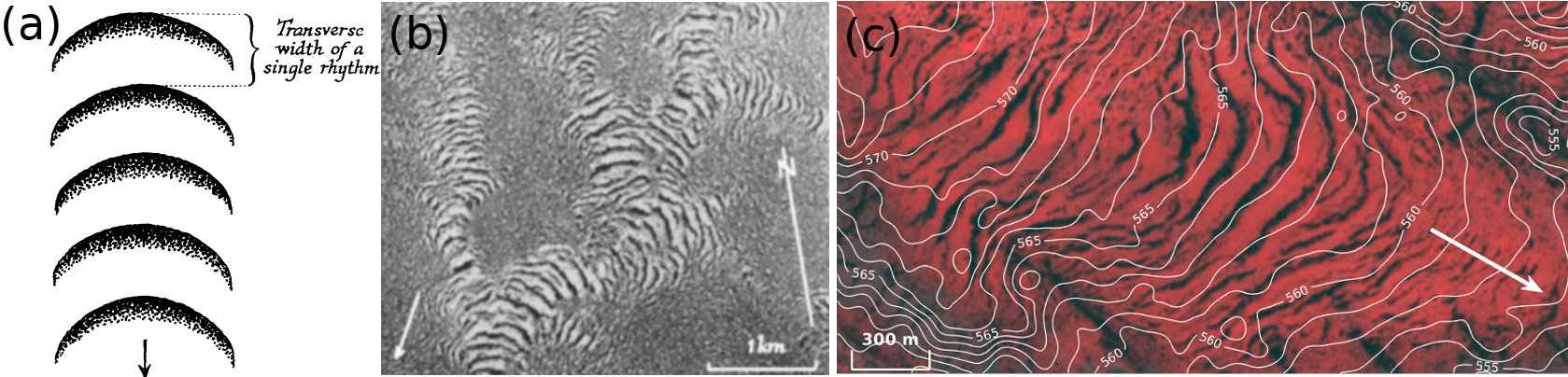}
    \caption{(a) Schematic of vegetation segments that are arced convex uplsope, reproduced from~\cite{macfadyen1950vegetation}. (b) Example of vegetation arcs confined to topographic depressions and oriented convex uplsope, also reproduced from~\cite{macfadyen1950vegetation}.  (c) Example of banded vegetation bands that are arced convex uplsope in ``valleys" and convex downslope on ``ridges," from Western Australia~\citep{gandhi2018topographic}.  The white arrows in (b,c) indicate the prevailing downhill direction. }
    \label{fig:arcing}
\end{figure}

The earliest observational study from the Horn of Africa~\citep{macfadyen1950vegetation}, noted that the vegetation bands tend to align transverse to the prevailing elevation gradient and arc convex upslope (Figure~\ref{fig:arcing}a,b). Later studies~\citep{boaler1964observations,greenwood1957development} point to examples where the vegetation bands were arced convex downslope and suggest that the arcing direction depends on the location of the patterns relative to cross-slope curvature in the terrain: the bands tend to follow elevation contour lines (Figure~\ref{fig:arcing}c).  Modern observations~\citep{penny2013local} support this claim that the bands tend to follow contours, but find that small-scale disruptions in topography or water flow can lead to significant deviation. The first modeling efforts that rely on ideas from pattern formation~\citep{lefever1997origin}, 
suggested that arced vegetation segments can form spontaneously on a uniformly sloped hill (with straight elevation contour line);  similar emergent arcing on a uniform slope was found in other models as well~\citep{meron2007localized,siero2015striped}.

Within the context of conceptual reaction-advection-diffusion pattern formation models,
\cite{gandhi2018topographic} made predictions about arcing direction by introduce a general topography $Z=\zeta(X,Y)$. They replaced the transport term $V\partial_X W$ in~\eqref{eq:klaus:w} with one based on a topography-dependent flux,
\begin{equation}
    \nabla\cdot( W\nabla\zeta)\equiv \nabla\zeta\cdot\nabla W+(\nabla^2\zeta)W,
    \label{eq:terrain}
\end{equation}
in order to investigate the impact of a cross-slope curvature on patterns. From~\eqref{eq:terrain}, we see that the transport leads to both an advective contribution ($\nabla\zeta\cdot\nabla W$), and a change in the (local) water loss rate  proportional to the terrain curvature  ($(\nabla^2\zeta) W$).
On ``ridge lines", where $\nabla^2\zeta<0$,  loss is increased due to water diversion, while in ``valley lines", where $\nabla^2\zeta>0$, loss is decreased due to water  accumulation.  The prediction from this model is that, as aridity increases, patterns shift from being predominantly on higher elevation ridge lines, convex downslope, to being spatially confined to lower elevation regions, convex upslope (as we see in Figure~\ref{fig:arcing}b).  
These model predictions find some traction when comparing arcing direction of vegetation bands  in different regions that have different aridity levels~\citep{gandhi2018topographic}. 
This modeling framework makes another, as yet untested, prediction: all bands migrate uphill, regardless of their curvature, and low-lying valley patterns with the same band spacing, are predicted to migrate faster than those on higher ground, 
all else being equal.

\section{Outlook}
\label{sec:outlook}

People study patterns because they are beautiful, but also because the semiarid regions where patterns occur sustain human life.  There may be information about the fragility of the system embedded in the patterns. Studying the patterning and vegetation patchiness that arises in these areas may also offer insights into the efficient use of resources in these water-limited places. Changes in vegetation cover with changing climate also have the potential to feedback into the climate system as a forcing \cite{rietkerk2011local, boussetta2015assimilation}.  Understanding land use, land use change, and vegetation evolution in patterned regions thus has implications for overall modeling of the climate system.

Climate change is predicted to pose a challenge in many drylands \citep{huang2017dryland}, expanding their extent and altering patterns of precipitation.  This has implications for how well drylands can continue to support populations that depend upon them.  Climate change projections for the Sahel and the Horn of Africa contain uncertainty, with models predicting different signs to rainfall change in some locations \citep{rowell2016can}.

\begin{figure}
    \centering
    \includegraphics[width=\linewidth]{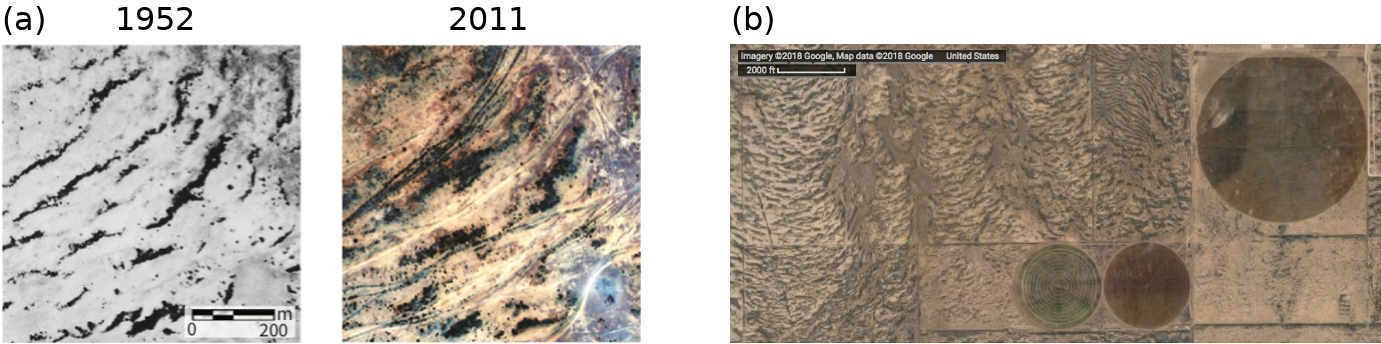}
    \caption{(a) An example of band degradation amid dense track cover in Somalia between 11/29/1952 and 12/03/2011 (9.58$^\circ$N, 48.57$^\circ$E). Figure is from \cite{gowda2018signatures} with images courtesy of the Bodleian Library and the DigitalGlobe Foundation.  (b) An example of agricultural land use in close proximity to vegetation patterns near Fort Stockton, Texas  (31$^\circ$02'58.6''N 103$^\circ$20'30.2''W).  Approximately 17 km by 7 km. Image from Google.}
    \label{fig:outlook}
\end{figure}

Human interactions with the ecosystem include, but are not limited to, direct harvesting, to impacting grazing pressure, and to land use change.  Other impacts (such as disrupting soil crust) have a number of human-driven causes \citep{belnap2001disturbance} and may be more difficult to parameterize in a model.  An example of this disruption can be seen in the extensive tracking shown in Figure \ref{fig:outlook}a.  This figure, taken from 
\cite{gowda2018signatures}, compares bands from 1952 and 2011.  In the 2011 image, tracks are likely from vehicles.  This soil disruption has spread vegetation out on the landscape and interrupted stripes.   Human actions can also generate regions of fertility in these systems.  For instance, creating patches has been used for restoration in a semiarid area \citep{ludwig1996rehabilitation}.  

Land use choices also impact the extent of land that may exhibit patterns.   
For example, in Figure \ref{fig:outlook}b, near Fort Stockton, Texas in the United States, rain-fed vegetation patterns are interrupted by irrigation driven agriculture.  The circular regions that are more highly vegetated are due to irrigation, while the banded patterns nearby are the local vegetation cover.  There has also been an expansion of cropland in the Sahel over the satellite era \citep{hiernaux2016desertification}. Changing land use, perhaps evolving with population pressure, sits outside the scope of current modeling efforts for vegetation patterns.

There is a disconnect between the reality of human-impacted systems, and the types of idealized, ecosystem-only models, that are straightforward to analyze.  Human pressures on biomass, on soil, and on land use, are difficult to parameterize in models.  In addition, human impacts have downstream effects on human livelihoods, so there is likely a feedback between human pressure and biomass productivity that is not captured in modeling work that neglects humans.  In the future, data driven modeling approaches combined with high-throughput remote sensing data have the potential to change how we frame this problem.  Observation-based work intrinsically includes human interactions with the system.  The growing back-and-forth between observations and modeling makes this a rich area of study.

\vspace{10mm}
\noindent\textbf{Acknowledgements:} We are grateful to Karna Gowda for insightful discussions and, along with Jake Ramthun, for help with figures.  This work was supported in part by the National Science Foundation grant DMS-1440386 to the Mathematical Biosciences Institute (PG) and National Science Foundation grant DMS-1517416 (MS). 

\begin{small}
\bibliographystyle{plainnat}
\bibliography{arxiv}
\end{small}

\end{document}